\documentclass{aa}

\usepackage{MR_AA}
\usepackage[dvips]{graphicx}
\usepackage{txfonts}
\usepackage{natbib}
\usepackage[T1]{fontenc}
\usepackage{subfigure}
\usepackage{spacing}

\newcommand{\csd}{C_{\rm sd}}
\begin{document}

\title{Dynamics of the radiative envelope of rapidly rotating stars:
Effects of spin-down driven by mass loss}
\titlerunning{Dynamics of the radiative envelope of rapidly rotating
stars}

\subtitle{}

\author{M. Rieutord\inst{1,2}
\and A. Beth\inst{1,2}} 

\institute{Universit\'e de Toulouse; UPS-OMP; IRAP; Toulouse, France
\and CNRS; IRAP; 14, avenue Edouard Belin, F-31400 Toulouse, France}

\date{\today}

\abstract
{}{This paper aims at deciphering the dynamics of the envelope of
a rotating star when some angular momentum loss due to mass loss is
present. We especially wish to know when the spin-down flow forced by
the mass loss supersedes the baroclinic flows that pervade the radiative
envelope of rotating stars.
}{
We consider a Boussinesq fluid enclosed in a rigid sphere whose flows are
forced both by the baroclinic torque, the spin-down of an outer layer,
and an outward mass flux. The spin-down forcing is idealized in two ways:
either by a rigid layer that imposes its spinning down velocity at some
interface or by a turbulent layer that imposes a stress at this same
interface to the interior of the star.
}{
In the case where the layer is rigid and imposes its velocity, we find
that, as the mass-loss rate increases, the flow inside the star shows two
transitions: the meridional circulation associated with baroclinic flows
is first replaced by its spin-down counterpart, while at much stronger
mass-loss rates the baroclinic differential rotation is superseded
by the spin-down differential rotation. When boundary conditions specify the stress
instead of the velocity, we find just one transition as the mass-loss
rate increases.  Besides the two foregoing transitions, we find a
third transition that separates an angular momentum flux dominated by
stresses from an angular momentum flux dominated by advection.  Thus,
with this very simplified two-dimensional stellar model, we find three
wind regimes: weak (or no wind), moderate, and strong. In the weak wind
case, the flow in the radiative envelope is of baroclinic origin. In
the moderate case, the circulation results from the spin-down while the
differential rotation may either be of baroclinic or of spin-down origin,
depending on the boundary conditions or more generally on the coupling
between mass and angular momentum losses. For fast rotating stars, our
model says that the moderate wind regime starts when mass loss is higher
than $\sim10^{-11}$\msun/yr. In the strong wind case, the flow in the
radiative envelope is mainly driven by angular momentum advection. This
latter transition mass-loss rate depends on the mass and the rotation
rate of the star, being around 10$^{-8}$\msun/yr for a 3~\msun\ ZAMS
star rotating at 200~km/s according to our model.
}{
}

\keywords{stars: atmosphere - stars: rotation}

\maketitle

\section{Introduction}

One of the well-known properties of rotating stars is their ability to
allow matter and angular momentum to be transported across their radiative
zone. Indeed, unlike non-rotating stars, a rotating, stably stratified
radiative zone cannot be at rest in any (rotating) frame. This property
was pointed out long ago by \cite{vz24}, but refer to \cite{R06b} for a
recent presentation. The mixing induced by rotation,
known as rotational mixing, is usually invoked to explain some features
of surface abundances in stars (Li-depletion or Nitrogen enrichment for
instance). Just as the rotation itself, however, variations of rotation are
also known to be an important source of mixing. The spin-down associated
with angular momentum loss, itself a consequence of a stellar wind,
generates a strong meridional circulation known as Ekman circulation
\cite[e.g.][]{R92,zahn92,RZ97}.

These features of the dynamics of rotating stars have been included in
stellar models using various recipes and assumptions. The main difficulty
is that the flows are essentially two-dimensional and therefore can
hardly be cast into a one-dimensional model. When this is done, the
azimuthal component of the flow field is the only remaining part of
the velocity field, which reads $\vv=r\Omega(r)\sth\ephi$. This is
the so-called ``shellular" differential rotation. The meridional flows
cannot be computed as such, since they are intrinsically 2D. Early work
\cite[e.g.][]{pinsonneault97} modelled transport as a diffusive process,
but \cite{zahn92} proposed another approach, now quite popular, which
takes the advective process of a meridional circulation into account.
Two-dimensional quantities are expanded in spherical harmonics and
averaged over isobars. Provided that horizontal diffusion dominates
vertical transport, advection of chemical elements by a meridional
circulation can be incorporated into a vertical effective diffusion.

The main difficulty with 1D models is that they only apply to slowly
rotating stars, since the spherical harmonics series is usually
severely truncated (up to $\ell=2$), which produces a poor representation
of the Coriolis effect \cite[][]{R87}. Actually, the trouble is that
we do not know the limiting rotation rate for a reliable use of 1D
models making  hazardous the use of these models for rapidly rotating stars,
or stars that have been rapidly rotating.  Thus, even if 1D
models have been appropriate guides in the interpretations of abundances
observations, a complete understanding of the effects of rotation is
still missing.

To go beyond one-dimensional models, we need to study the flows that
take place in rotating stars so as to understand their dependence with
respect to the main features of stellar conditions (\BVF\ profiles,
turbulence, thermal diffusivity, etc.). This kind of study was initiated
by \cite{R06} or \cite{ELR07} with no account of a possible angular
momentum loss of the star, however, the spin-down induced by such a
process is likely to be a very important part of the dynamics of massive
stars \cite[e.g.][]{Lau_etal11} or of young stars \cite[][]{LCM96},
and therefore deserves a detailed study.

In order to progress in the understanding of the dynamics
in these stars, we investigate the effect of the spin-down using a
Boussinesq model of a star, thus completing the work of \cite{R06}.
Although such a model is quite unrealistic as far as direct comparisons
to observational data are concerned, it is a useful step to
enlight the  basic mechanisms operating in a spinning down mass-losing
star, and to later deal with more realistic models.

With such a model, we wish to determine the relative influence
of baroclinicity and Ekman circulation associated with spin-down
on meridional advective transport, and appreciate, when possible,
the dynamical consequences of a given mass-loss rate. A review of
spin-up/spin-down flows, from the viewpoint of fluid dynamics and
including somehow the effect of stratification, may be found in
\cite{duck_etal01}.

The paper is organized as follows: in the next section, we detail
the model that we are using, especially the physics that is included.
In the following section, we analyse the spin-down flow when it is either
driven by an imposed velocity field at the top of the stellar interior,
or when it is driven by stresses imposed by the spinning-down layer where
the stellar wind is rooted. The results of the foregoing fluid dynamics
section are then discussed in the astrophysical context in term of the
wind strength (Sect.~4). We then summarize the hypothesis and results
of this work in the last section. The hurried reader solely interested
in the astrophysical conclusions may jump directly to this final section.

\begin{figure}
  \begin{center}
   \includegraphics[scale=0.2]{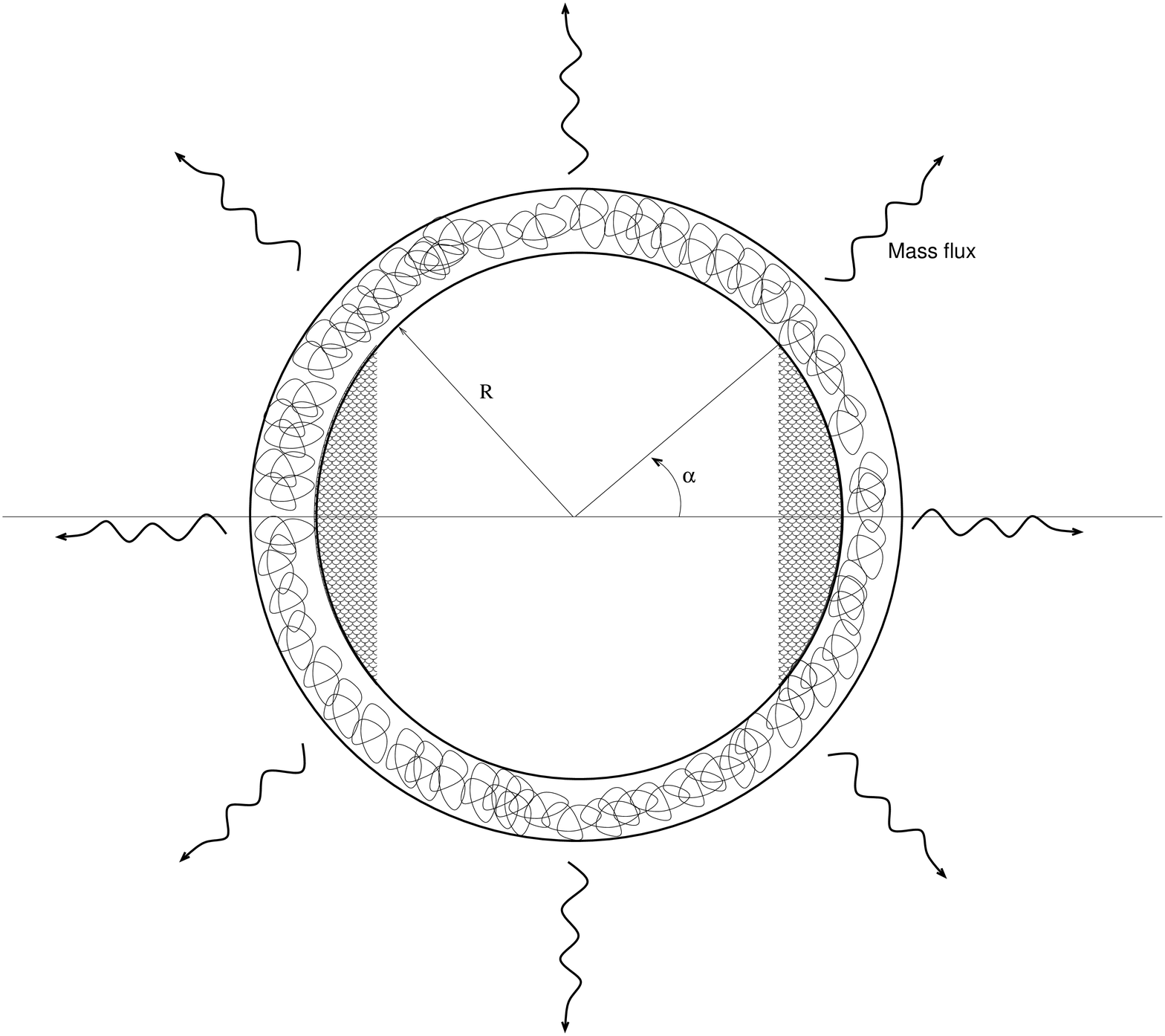}
\caption{Schematic view of the system: the spinning down turbulent
envelope surrounds stably stratified fluid where the spin-down
flow develops. The scale-filled outer cylinder of fractional radius
s=$\cos\alpha=2/\sqrt{7}$ is an unstable region that may exist in some
cases (see \S~\ref{centr_inst}).}
\label{croquis}
\end{center}
\end{figure}

\section{The model}

\subsection{Description}

We consider a self-gravitating, viscous fluid of almost constant
density, enclosed in a spherical box of radius $R$. With this
boundary we try to mimic an upper turbulent boundary layer, likely
threaded by magnetic fields, which is rotating rigidly or differentially
and thickening with time. 

The dynamical interaction of a stellar wind with the stellar interior is
far from being fully understood. Most studies rely on a global balance of
angular momentum \cite[e.g.][]{zahn92,Lau_etal11}. Following \cite{LCM00},
we imagine that the friction between the angular momentum losing layer
makes it turbulent and that this turbulence entrains lower layers, by the
well-known turbulent entrainment process \cite[][]{Turn86}. The turbulent
layer thus slowly deepens while extracting angular momentum from the
star's interior, however, at the same time the star slowly expands as
mass is removed, and some outward radial flow also contributes to the
spin-down process.

To mimic this complex phenomenon, we assume that the turbulent layer is
like a highly viscous fluid that is absorbing some mass flux from the
interior. At the interface,  the conditions met by the velocity field
demands the continuity of both the velocity field and the associated
stresses. Since these conditions are quite involved (they need the
mean flow field in the turbulent layer), we shall reduce them to two
ideal cases: (i) The turbulent layer rotates rigidly and therefore
imposes a solid body rotation at the interface, whose angular velocity
evolves as $\Omega = \Omega_0 +\dot{\Omega}t$, with $\dot{\Omega}<0$
and $|\dot{\Omega}t|\ll \Omega_0$. (ii) The turbulent layer imposes a
stress, which brakes the fluid below.  In both cases, however, some mass
flow crosses the boundary.

Such a modelling is inspired from the work of \cite{fried76} who considered
a similar configuration of a Boussinesq stably stratified fluid inside a
sphere, which experiences spin-down by a given surface stress. She discussed this problem
using linearized equations thus considering a weak spin-down. One result
of this work is that the solutions of this linear problem split into two
components: first, a component growing linearly with time and identifiable to
a solid body rotation (the actual spinning down rotation) and second,
steady, a component made of a meridional circulation that carries the angular
momentum, and a differential rotation. As a result, \cite{fried76}
could relate the torque imposed by the surface stress and the angular
deceleration of the fluid.

Our rigid condition (i) therefore extends this previous work, but, as we
shall see, many features of the solutions are common to the two cases.
We sketch out this model in Fig.\ref{croquis}.

\subsection{Equations of motion}

\begin{figure}
  \begin{center}
   \includegraphics[scale=0.22]{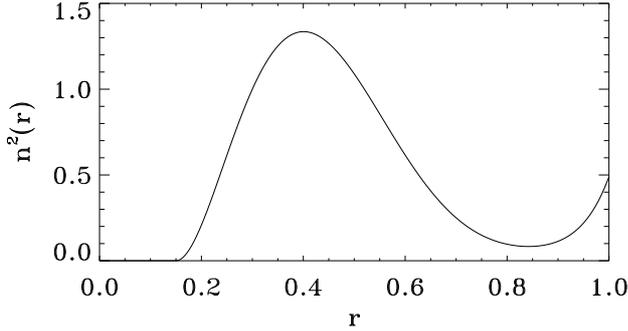}
\caption{Adopted profile for the Brunt-V\"ais\"al\"a frequency in our
calculations. We take the core radius at $r=0.15$}
\label{bvfp}
\end{center}
\end{figure}

\begin{table*}
  \begin{center}
\begin{tabular}{cccccccccccc}
\hline
&&&&&&&&&\\
Mass & Radius & P$_{\rm break-up}$&$\moym{\mathcal{N}^2}$&
$\moym{\nu}$ & $\moym{\PR}$& $\nu_t$ & E$_t$&
$\lambda_{\rm fast}$& $\lambda_{\rm slow}$& V$_{\rm eq}^{\rm fast}$& V$_{\rm eq}^{\rm
slow}$\\
(M$_\odot$)&(\Rsun)&(days)&(s$^{-2}$)&(cm$^2$s$^{-1}$)&&(cm$^2$s$^{-1}$)&&&&(km/s)&(km/s)\\
&&&&&&&&&&&\\
3   & 1.96 & 0.28 & $2.6\times10^{-6}$ & $3\times10^{6}$ & $10^{-6}$&
$5\times10^9$ &
$8\times 10^{-10}$ & $5.6\times10^{-5}$& 0.29 & 200 & 2.8 \\
&&&&&&&&&&&\\
7   & 3.15 & 0.41 & $1.3\times10^{-6}$ & $3\times10^{7}$ & $9\times10^{-7}$&
$1\times10^{11}$ &
$9\times10^{-9}$ & $1.8\times10^{-5}$ & $9\,10^{-2}$ & 320 & 4.4 \\
&&&&&&&&&&&\\
\hline
\end{tabular}
  \end{center}
\caption[]{Parameters of two intermediate-mass ZAMS stars. The $\lambda_{\rm
fast}$ and $\lambda_{\rm slow}$ parameters refer to the fast rotating
star (rotation period of 0.5 day) and the slow rotating star (rotation
period of 36 days).}
\label{param_stars}
\end{table*}

The gravity field inside the fluid is simply $\vec{g}=-g_s \vec{r}$
where $g_s$ is the surface gravity and $\vec{r}$ is the reduced
radial coordinate (i.e $r=0$ at the centre and $r=1$ at the outer
boundary).
At equilibrium, the fluid is governed by:

\begin{equation}
\left \{
\begin{array}{l}
-\vec{\nabla}P_{eq}+\rho_{eq}\vec{g}=\vec{0}\\
\vec{\nabla}.(\chi\vec{\nabla}T_{eq})+Q=0\\
\rho_{eq}=\rho_{0}(1-\alpha(T_{eq}-T_{0})) \\
\end{array}
\right .
\label{equilibre}
\end{equation}
where $\alpha$ is the dilation coefficient, $\chi$ the thermal
conductivity, and $Q$ the heat sinks (inserting heat sinks in the fluid
is a trick to impose a stable stratification). Here, $P_{eq}, \rho_{eq}$
and $T_{eq}$ are the equilibrium values of the pressure, density,
and temperature respectively. We shall need the Brunt-V\"ais\"al\"a
frequency, namely

\begin{equation}
N^{2}(r)=\alpha \frac{dT_{eq}}{dr}g(r)\; .
\label{brunt}
\end{equation}
As in \cite{R06}, we mimic the \BVF\ profile of stars with the
simplified profile shown in Fig.~\ref{bvfp}.

We now let the system rotate at an angular velocity $\Omega$ around the
$z$-axis but, contrary to \cite{R06}, we assume that $\Omega$ slowly
decreases with time, thus

\[ \dot{\Omega}=\frac{d \Omega}{dt} < 0\; .  \]
By slow we mean that $\dot{\Omega}/\Omega^2\ll1$, namely that the
rotation rates varies very little during a rotation period.
In the co-rotating frame, steady flows are the solution to the following
equations:

\begin{equation}
\begin{array}{rcl}
\Div\vv &=& 0\; ,\\
\rho\ (2\vec{\Omega}\wedge\vec{v} +
\vec{\dot{\Omega}}\wedge\vec{r} +
(\vec{v}\cdot\vec{\nabla})\vec{v}\ )
&=& -\vec{\nabla}P +
\rho(\vec{g}+\Omega^{2}s\vec{e}_{s}) +
\mu\Delta\vec{v}\; ,\\\\
\rho c_v(\vec{v}\cdot\vec{\nabla}T)&=&\vec{\nabla}\cdot
(\chi\vec{\nabla}T)+Q \; ,
\end{array}
\label{eq3}
\end{equation}
which express the conservations of mass, momentum, and energy,
respectively. There, $\mu$ is the dynamical shear viscosity, $c_v$ the
specific heat capacity at constant volume, $s$ the radial cylindrical
coordinate, and $\vec{e}_s$ the associated unit vector. These equations
differ from those of \cite{R06} by the new term
$\vec{\dot{\Omega}}\wedge\vec{r}$, also called the Euler acceleration.
Introducing fluctuations with respect to the equilibrium set-up
described by \eq{equilibre} and following \cite{R06} we derive the
vorticity equation, which we complete with the equations of energy and mass
conservation, namely

\begin{equation}
\left \{
\begin{array}{l}
\vec{\nabla}\wedge
\lc 2\vec{\Omega}\wedge\vec{v}+(\vec{v}\cdot\vec{\nabla})\vec{v}
+\alpha\delta T(\vec{g}+\Omega^{2}s\vec{e}_{s})-\nu\Delta\vec{v}\rc = \\
 \hspace*{3cm}-\epsilon
N^{2}(r)\sin\theta\cos\theta\vec{e}_{\phi}-2\dot{\Omega}\vec{e}_{
z}\\
\vec{v}\cdot\vec{\nabla}T_{eq}+\vec{v}\cdot\vec{\nabla}\delta T
=\kappa\Delta\delta T\\
\vec{\nabla}\cdot\vec{v}=0\; .
\end{array}
\right .
\end{equation}
Here, $\kappa$ is the thermal diffusivity, and \[ \epsilon=\Omega^2
R/g_s = \lp\frac{\Omega}{\Omega_k}\rp^2\] is the ratio of centrifugal
acceleration to surface gravity, $\Omega_k$ being the associated keplerian
angular velocity.

\subsection{Scaled equations}

Our problem is forced. We need now to scale these equations
to get solutions of order unity. Thus, we set

\begin{equation}
\vec{v}=\frac{\epsilon \mathcal{N}^2 R}{2\Omega}\vec{u},
\quad \delta T=\epsilon T_{*}\vartheta, \quad
\mathcal{N}^2=\frac{\alpha T_{*} g_s}{R}
\end{equation}
where $\mathcal{N}$ is the scale of the Brunt-V\"ais\"al\"a frequency.
Finally, we obtain the equations for dimensionless dependent variables:

\begin{equation}
\left \{
\begin{array}{l}
\displaystyle{
\vec{\nabla}\wedge \left (
\ez\wedge\vu+\RO(\vu\cdot\na)\vu -\vartheta(r\er+\epsilon s\es)-E \Delta\vu \right)}=\\
\hspace*{3cm}\displaystyle{-
n^{2}(r)\sin\theta\cos\theta\vec{e}_{\phi}-2\csd\ez
}\\
\displaystyle{
\frac{n^{2}(r)}{r} u_r+\epsilon\
\vec{u}\cdot\vec{\nabla}\vartheta=\tilde{E}_T\Delta \vartheta
}\\
\displaystyle{
\vec{\nabla}\cdot\vec{u}=0
}
\end{array}
\right .
\label{eqstage1}
\end{equation}
where $n^2(r)$ is the scaled Brunt-V\"ais\"al\"a frequency, and

\begin{eqnarray}
&&E=\frac{\nu}{2\Omega R^2},\qquad \tE_T=\frac{\kappa}{2\Omega
R^2} \left ( \frac{2\Omega}{\mathcal{N}}\right )^2,\\
&&\RO=\epsilon\frac{\mathcal{N}^2}{4 \Omega^2}, \qquad {\rm and}\quad
\csd = \frac{\dot{\Omega}}{\epsilon \mathcal{N}^{2}}
\end{eqnarray}
where $E$ is the Ekman number, which measures the ratio of the viscous force
to the Coriolis force, $\tilde{E}_T$ measures heat diffusion, $\RO$
is the Rossby number and $\csd$ is the non-dimensional torque density
due to spin-down. For later use we also introduce

\begin{equation}
\mathcal{P}=\frac{\nu}{\kappa},
\quad \lambda=\frac{E}{\tE_T}=\frac{\mathcal{P}\mathcal{N}^2}{4\Omega^2}
\end{equation}
namely, the Prandtl number and its product with the scaled \BVF.

\subsection{Boundary conditions}

System \eq{eqstage1} needs to be completed by boundary conditions.

In the first case (i) where the spinning-down layer rotates rigidly and since
we are using a frame co-rotating with this layer, the continuity of
the velocity field at the interface of the turbulent layer leads
to the no-slip conditions\footnote{No-slip or rigid boundary
conditions assume that on the boundary the fluid has the same velocity
of the (assumed) solid wall that bounds it.} on the velocity
field. Looking for steady state solutions, we neglect the motion of the
interface but include a mass flux at the boundary as in \cite{HR13}.
Thus, we impose:

\begin{equation}
\vu=u_e\er\at r=1\; .
\label{bc0}
\end{equation}
In the second case (ii), we impose the tangential (azimuthal) component
of the stress, depending on co-latitude $\theta$, namely

\beq
\dr{}\lp\frac{u_\varphi}{r}\rp = -\tau(\theta), \quad
\dr{}\lp\frac{u_\theta}{r}\rp = 0\at r=1
\eeqn{bc1}
and

\beq u_r(1)=u_e \eeq
where we used non-dimensional quantities. The dimensionless stress
$\tau(\theta)$ is related to the dimensional stress $\tau_*(\theta)$ by

\beq \tau = \frac{2\Omega\tau_*}{\epsilon\calN^2\mu}\; .\eeqn{taund}
Here, we choose $\tau\geq0$ so that condition \eq{bc1} imposes the
braking of the fluid.

As far as the temperature field is concerned, we impose zero temperature
fluctuations on this surface, namely

\begin{equation}
\vartheta(1)=0\; .
\end{equation}
As shown by \cite{fried76}, this condition is of little importance
for the flow.

\subsection{Relation between spin-down and mass loss}

As mentioned above, the expansion of the envelope is also driving a
differential rotation and a meridional circulation.  To make things
tractable, we need to parametrize the mass flux while keeping its
association with the direct spin-down drivers (stress or velocity
conditions).

To this end, let $\dot J$ be the angular momentum loss of the star. This
flux is supplied at the base of the turbulent layer (where we set boundary
conditions) by a viscous torque and an angular momentum flux associated
with the outflowing mass. If we assume that the associated dimensional
viscous stress in \eq{bc1} is $\tau_*(\theta)=\tau_*\sth$ as suggested
by \cite{fried76}, the conservation of angular momentum flux across the
layer leads to

\begin{equation}
-\frac{8\pi}{3}R^3\tau_*+\frac{2}{3}\dotM R^2\Omega = \dotJ
\label{amb}
\end{equation}
where we assume that the rotation at the boundary is almost uniform. We
now  parametrize $\dotJ$ as $\beta\frac{2}{3}\dotM R^2\Omega$. The
total angular momentum flux in the layer is therefore split into a
fraction $\beta^{-1}$ of simple advection and $1-\beta^{-1}$ of viscous
stress. We thus write

\[ -\frac{8\pi}{3}R^3\tau_* = (\beta-1)\frac{2}{3}\dotM R^2\Omega\; .\]

Moving to non-dimensional quantities, the previous equation leads to

\beq u_e=\frac{2E\tau}{\beta-1} \eeq
which relates the expansion velocity and the stress.

In the case where spin-down is imposed by the velocity field (case (i)
of our boundary conditions), things are more involved because we still
need the stress to evaluate the angular momentum flux. Since the
associated torque is  due to the angular deceleration of the turbulent
layer, however, dimensional analysis leads to the following equation:

\beq k\rho\dotOm R^5 + \frac{2}{3}\dotM R^2\Omega = \dotJ\; ,\eeqn{defk}
where $k$ is a non-dimensional constant to be determined from the flow.
Introducing parameter $\beta$ as before, we have 

\[ \frac{\dotOm}{\Omega} = \frac{2(\beta-1)}{3k}\frac{\dotM}{\rho R^3}\;
.\]
Turning to non-dimensional quantities, we find that the
expansion velocity is related to the non-dimensional torque density of
the spin-down $\csd$ by

\beq u_e = -\frac{3k}{4\pi(\beta-1)}\csd \; .\eeq

\begin{figure}
\begin{center}
\includegraphics[scale=0.30]{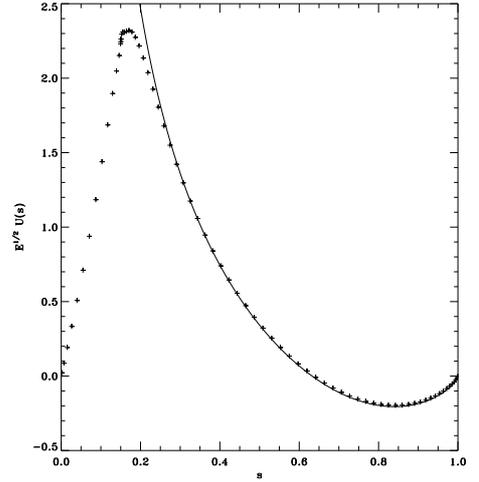}
\caption{Comparison between the analytical (solid line) and numerical
(+) solutions of the equatorial differential rotation of a
spin-down flow (E=$10^{-6}$ and $\beta=3$). $s$ is the radial coordinate.}
\label{u_geo}
\end{center}
\end{figure}

\subsection{Linearization}

We shall further simplify the problem by
letting $\dot{\Omega}\rightarrow0$ and $\epsilon \rightarrow 0$, but keeping
$\csd$  finite:

\begin{equation}
\left \{
\begin{array}{l}
\displaystyle{
\vec{\nabla}\wedge\lp\ez\wedge\vu- \vartheta r\er-E\Delta\vu\rp =
-n^{2}(r)\sth\cth\ephi-2\csd\ez
}\\\\
\displaystyle{
\frac{n^{2}(r)}{r} u_r=\tE_T\Delta \vartheta
}\\\\
\displaystyle{
\vec{\nabla}\cdot\vec{u}=0\; .
}
\end{array}
\right .
\label{eqstage}
\end{equation}
These PDE are completed by the following boundary conditions at $r=1$:

\[ u_r=u_e\]
and

\[ \dr{}\lp\frac{u_\theta}{r}\rp = 0\quad \&\quad
\dr{}\lp\frac{u_\varphi}{r}\rp = -\tau(\theta) \orr u_\theta=u_\varphi=0\]

We thus get a linear system where the velocity field results from the
superposition of three forcings:

\begin{itemize}
\item the baroclinic torque $-n^{2}(r)\sth\cth\ephi$,

\bigskip
\item \[ \left\{ \begin{array}[c]{l} {\displaystyle
\rm the\; spin-down\; (Euler)\; torque\; -2\csd\ez }\\ 
 \displaystyle{ \qquad\rm or}\\
 \textstyle{\rm the\; braking\; stress\; -\tau(\theta)}
\end{array}
\right.\]

\bigskip
\item the expansion flow
\end{itemize}

The effects of the baroclinic torque have been studied in \cite{R06}.
We thus focus on the spin-down part, which actually contains two
drivers: the friction between layers and the expansion flow.

\subsection{Typical numbers}

Before that and to fix ideas, we computed in Tab.~\ref{param_stars}
the main numbers for two ZAMS stars of intermediate masses. These
models of solar compositions have been computed with the TGEC
code (Toulouse-Geneva Evolution Code, see \citealt{hbh08}). From
these models, we estimate the typical values of the \BVF\ squared
$\moym{\mathcal{N}^2}$, the mean Prandtl number $\moym{\PR}$, the mean
kinematic (radiative) viscosity $\moym{\nu}$, the typical turbulent values
of the kinematic viscosity estimated from \cite{zahn92} (see below), and
the associated Ekman number. We shall also consider two typical rotation
periods, namely 0.5 day and 36 days, so as to represent a fast and
slow rotation.  These figures directly control the $\lambda$-parameter,
whose two extreme values are given. The rotation period also influences
the Ekman number, but in view of the uncertainties on the turbulent
transport, we prefer keeping a single value for this parameter. We also
give the break-up period below which the star loses mass at its equator
\cite[see][]{REL13,ELR13}.

\begin{figure}
\begin{center}
   \includegraphics[scale=0.4]{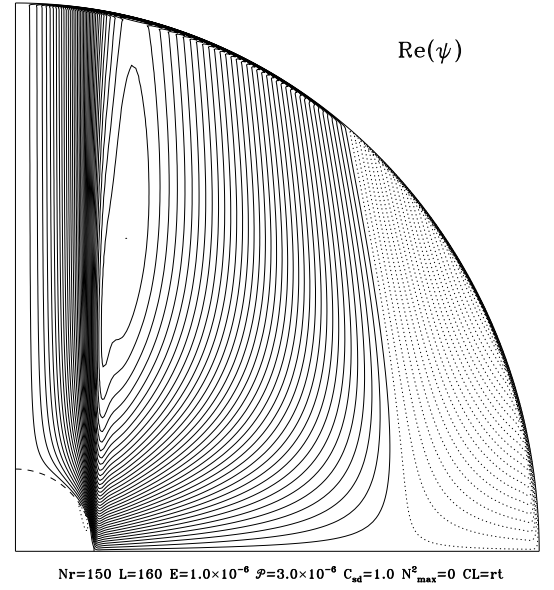}
\caption{Meridional circulation associated with a spin-down flow in the
case of Fig.~\ref{u_geo}. The dotted isocontours show a clockwise
circulation, while the solid lines are for anti-clockwise circulation.
Numerical resolution used Nr=150 Chebyshev polynomials radially and
spherical harmonics up to order L=160.
}
\label{u_merid}
\end{center}
\end{figure}

\section{The spin-down flow}

\subsection{Driven by velocity boundary conditions}\label{rigcase}

We first concentrate on the case where the outer turbulent layer spins
down as a solid body. In a frame co-rotating with this layer, the
non-dimensional velocity field $\vu$ is forced by the torque density
$-2\csd\ez$ and meets no-slip boundary conditions with an outflowing mass \eq{bc0}.

\begin{figure*}
  \begin{center}
\centerline{
\includegraphics[scale=0.3]{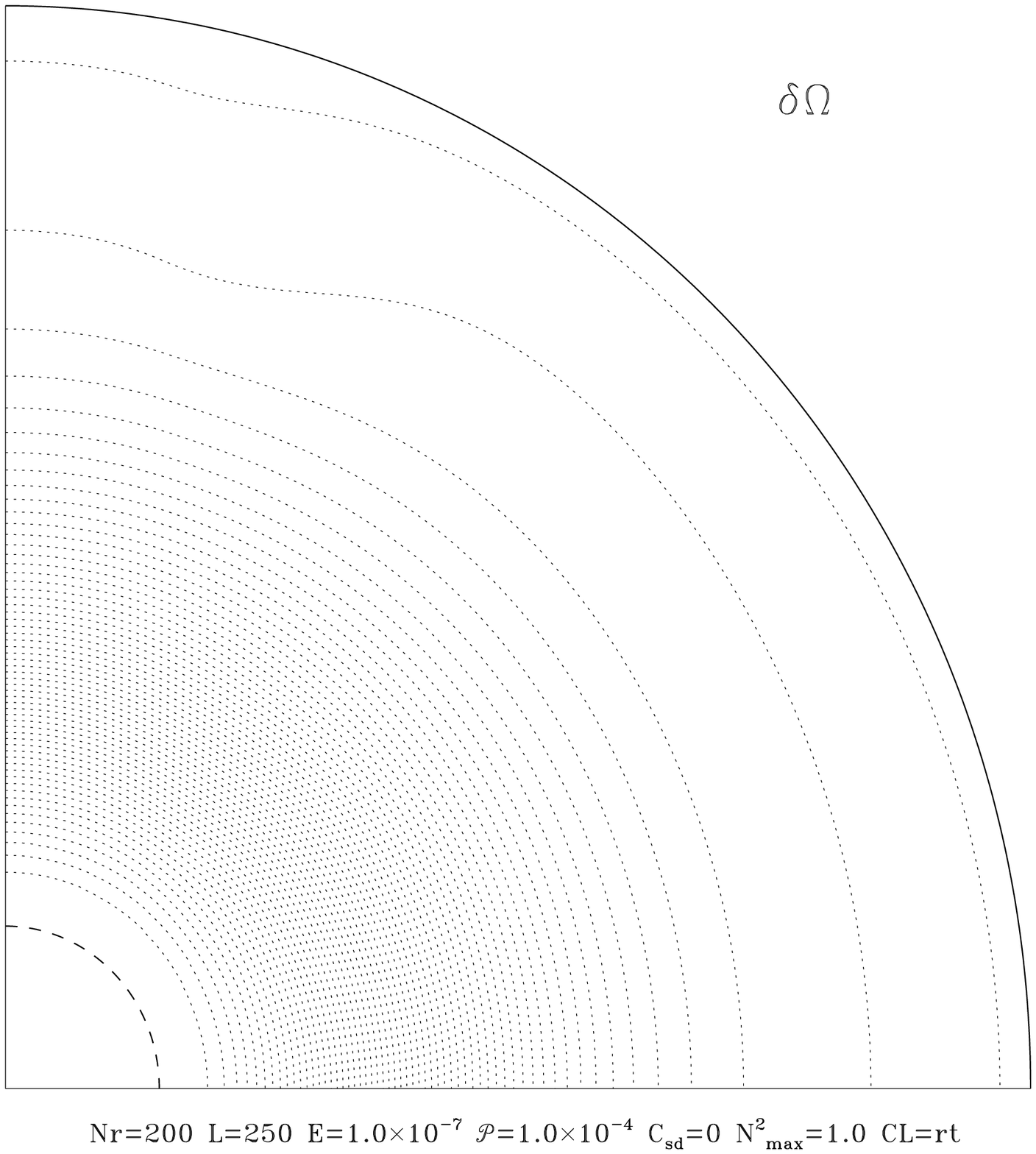}
\includegraphics[scale=0.3]{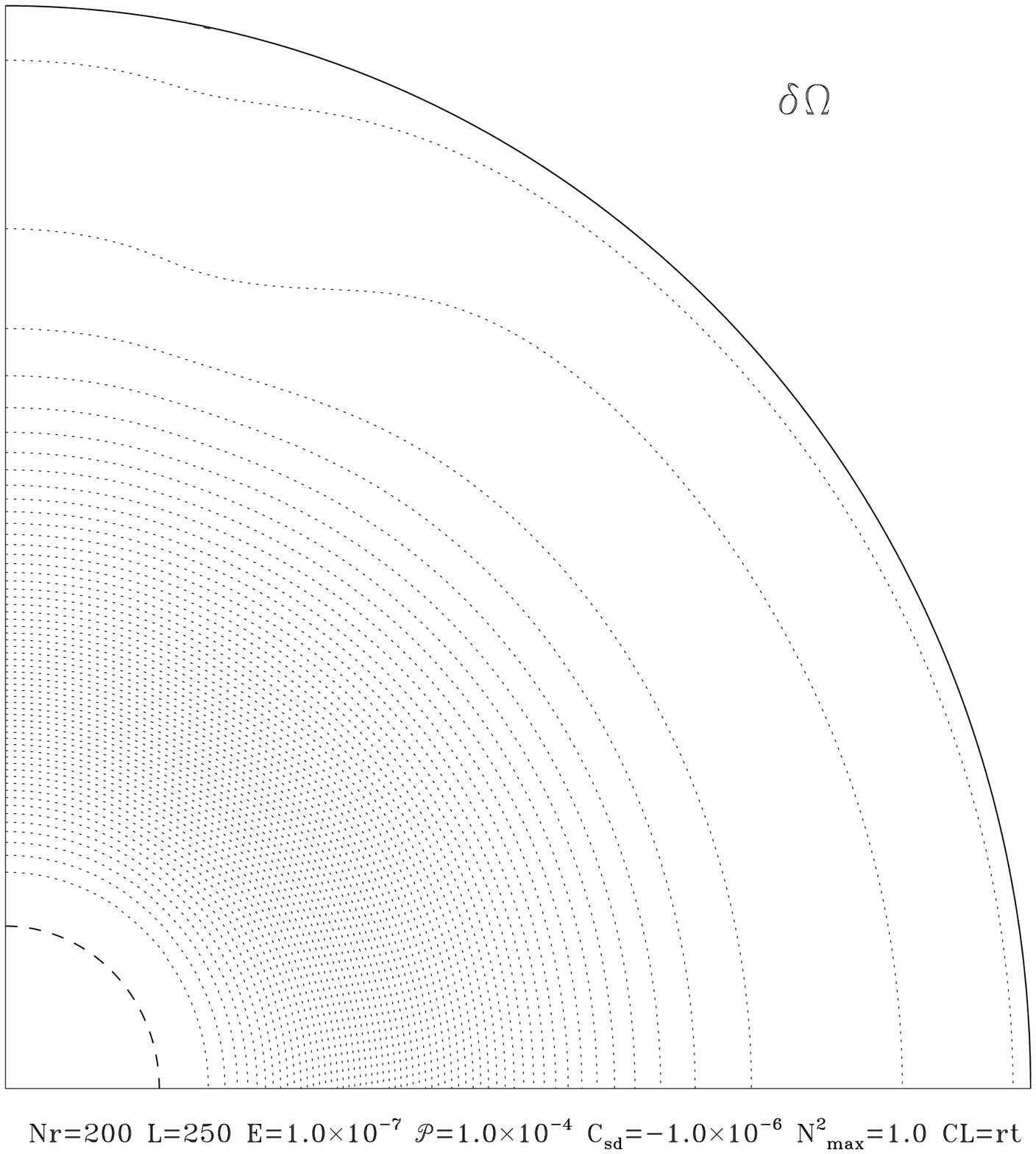}
\includegraphics[scale=0.3]{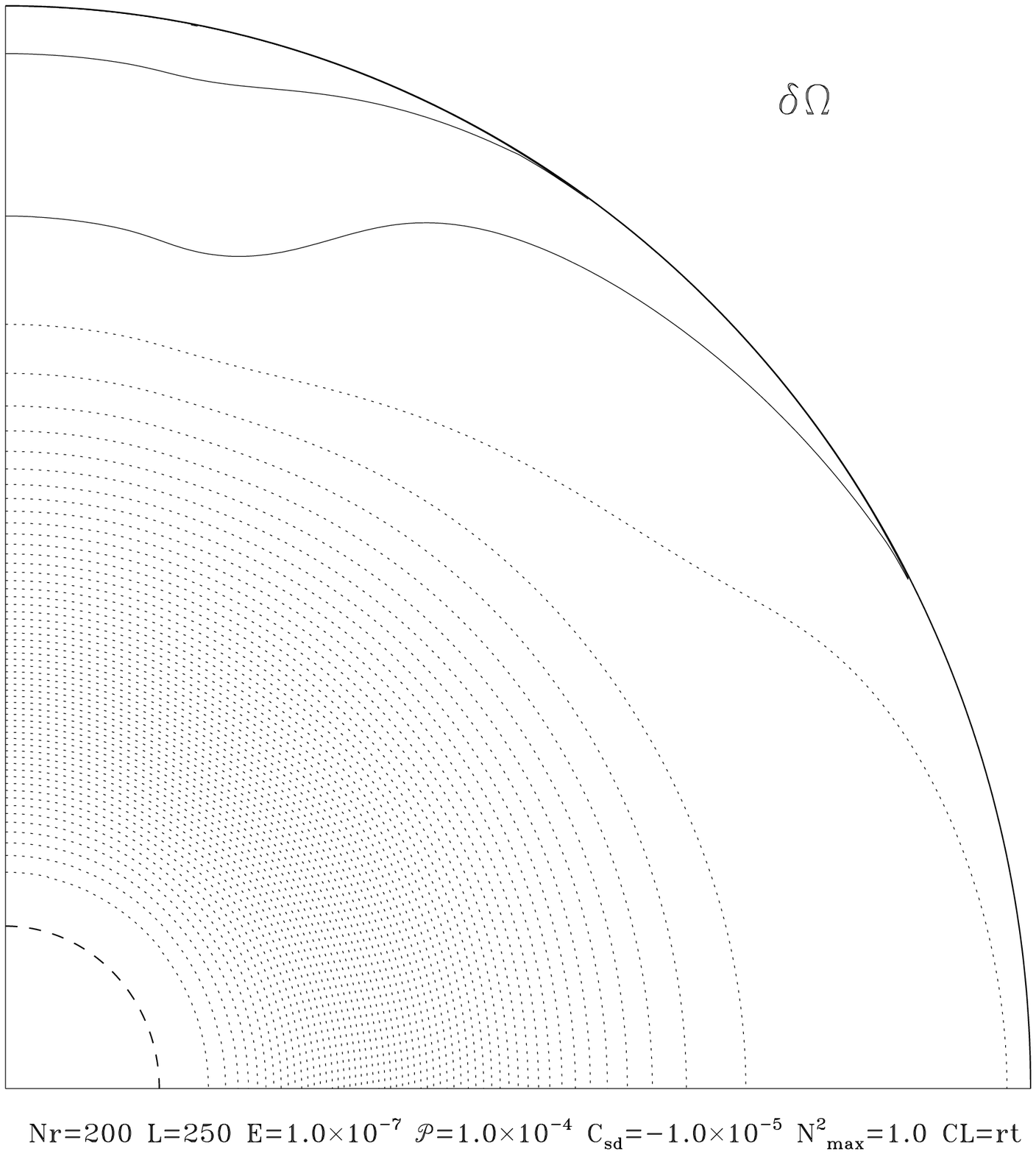}
}
\centerline{
\includegraphics[scale=0.3]{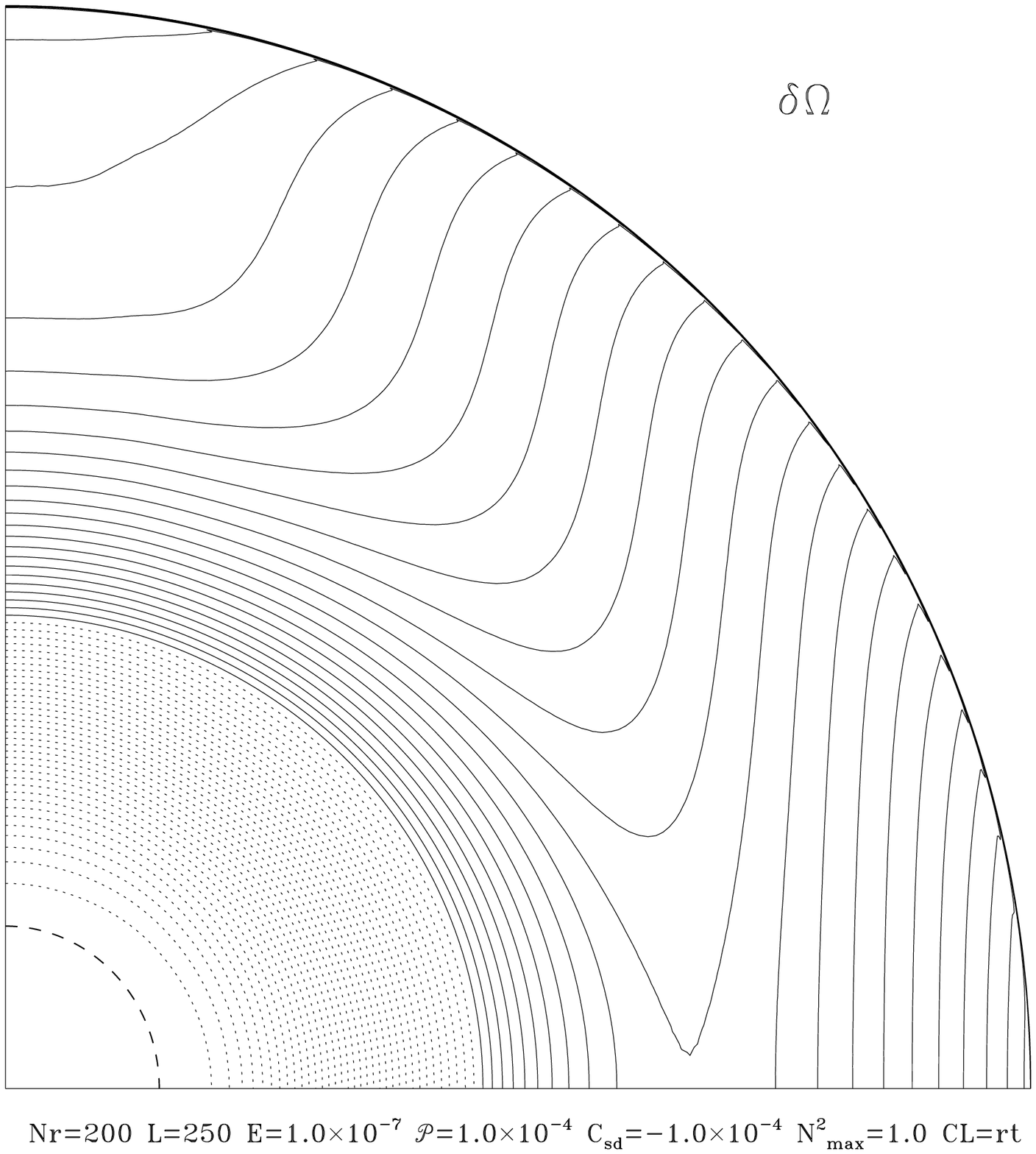}
\includegraphics[scale=0.3]{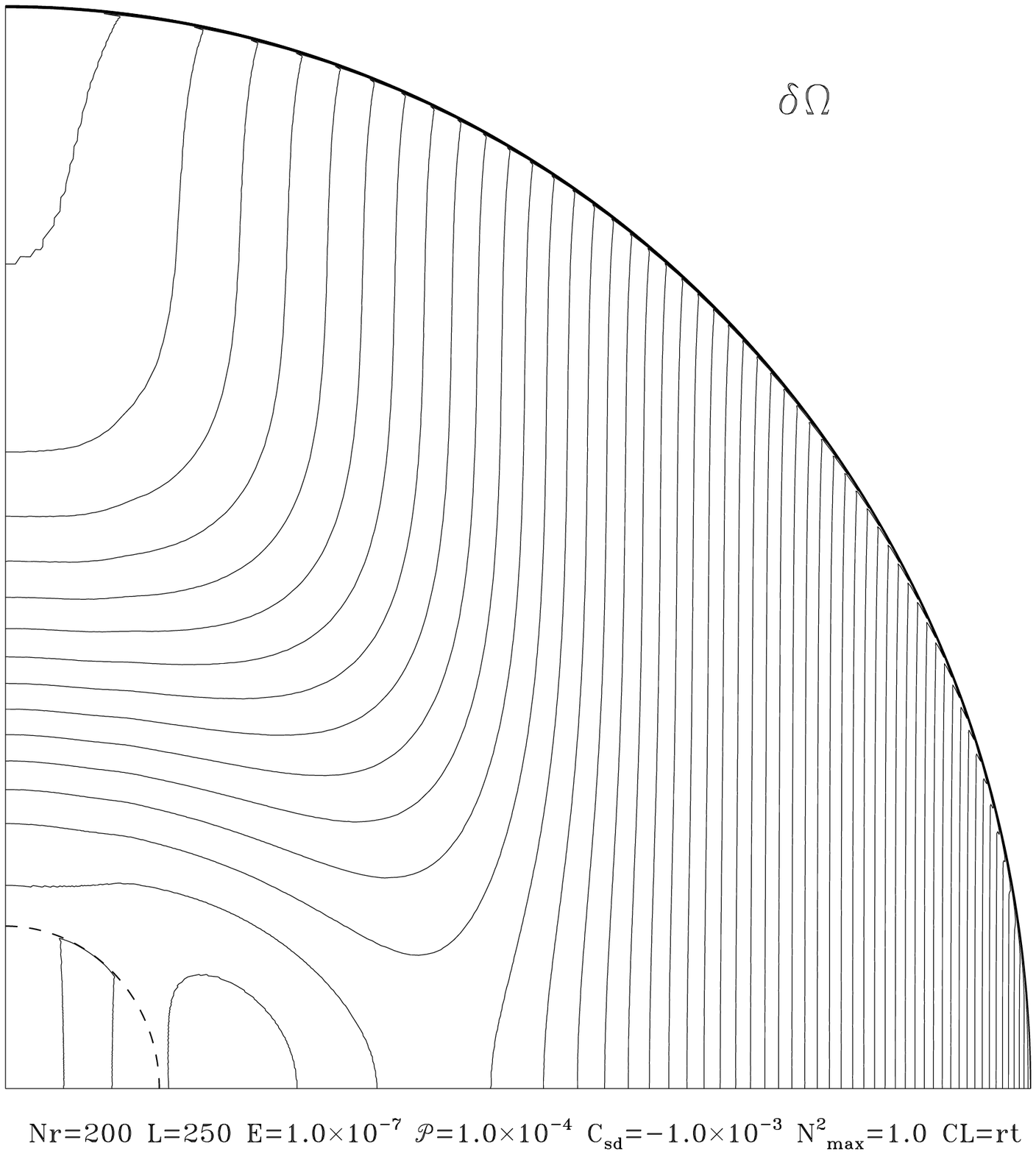}
\includegraphics[scale=0.3]{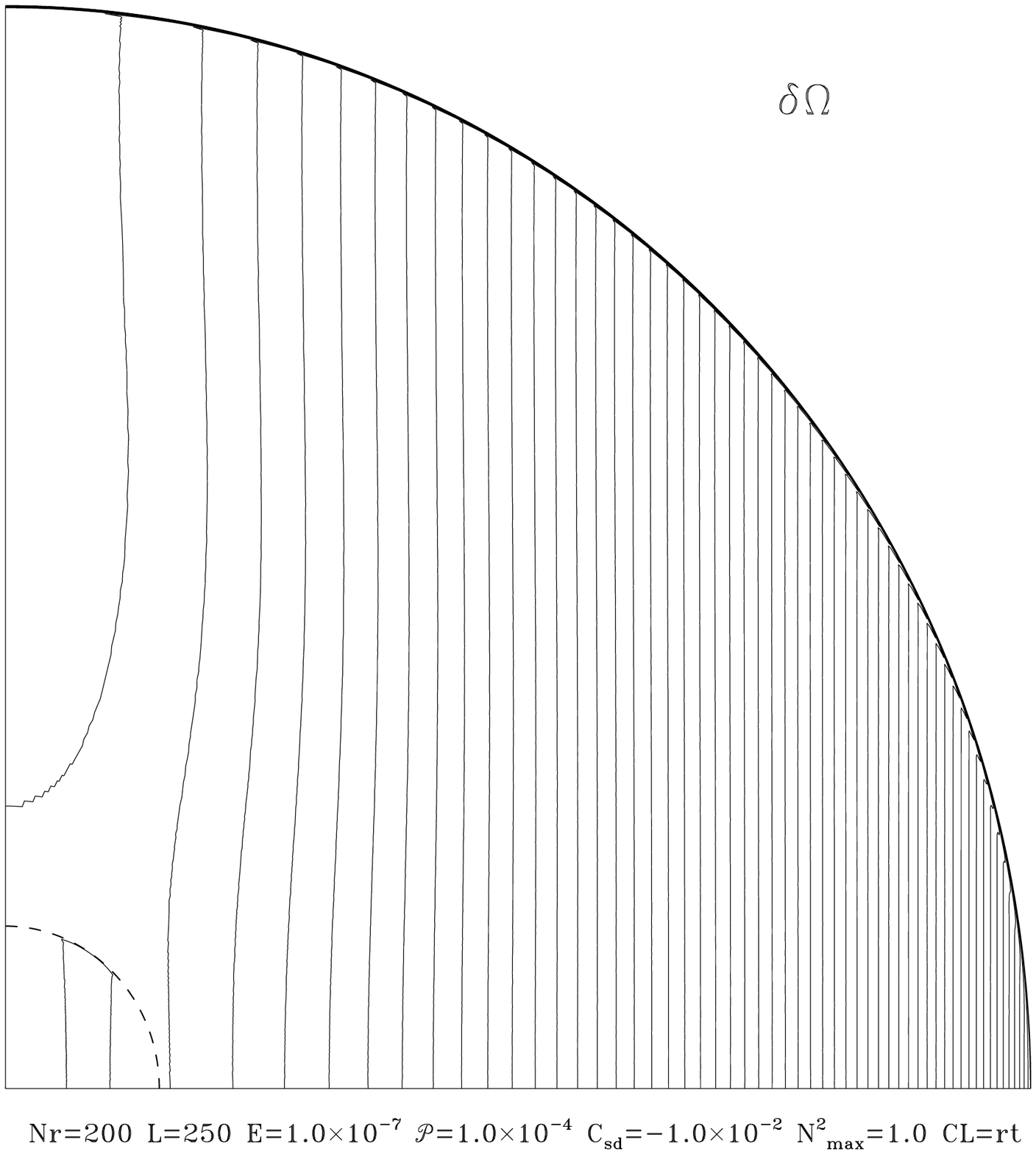}
}
  \end{center}
\caption{
Differential rotation of a stably stratified rotating fluid in a
sphere when the spin-down forcing $\csd$ is
strengthened. From left to right and top to bottom $\csd=0, -10^{-6},
-10^{-5}, -10^{-4}, -10^{-3}, -10^{-2}$. Ekman and Prandtl numbers are
$E=10^{-7}$, $Pr=10^{-4}$. All calculations have assumed $\beta\gg1$.
Solid lines are for positive values, dotted for negative values.
Numerical resolution used Nr=200 Chebyshev polynomials radially and
spherical harmonics up to order L=250.
}
\label{transit}
\end{figure*}

\subsubsection{Case of a negligible buoyancy}

As a first step, we neglect the buoyancy term $\theta_Tr\er$ and
give below the circumstances in which it is indeed negligible. Thus,
we first solve:

\begin{equation}
\left \{
\begin{array}{l}
\displaystyle{
\vec{\nabla}\wedge \left ( \vec{e}_z\wedge\vec{u}-E\Delta\vu \rp= 
-2C_{sd}\ez
}\\\\
\displaystyle{
\vec{\nabla}\cdot\vec{u}=0\; .
}
\end{array}
\right .
\label{eqsd}
\end{equation}
System \eq{eqsd} may be solved by a boundary layer analysis in the limit
of small Ekman numbers following \cite{R87}. As in \cite{HR13},
we change the outflow driving of the boundary conditions into a volumic
force by setting

\[ \vu = \vu'+\frac{u_e}{r^2}\er\]
so that \eq{eqsd} now reads

\begin{equation}
\left \{
\begin{array}{l}
\displaystyle{
\vec{\nabla}\wedge \left ( \vec{e}_z\wedge\vec{u}'-E\Delta\vu' \rp= 
-2C_{sd}\ez-\frac{u_e}{r^2}(2\cth\er+\sth\etheta)
}\\\\
\displaystyle{
\vec{\nabla}\cdot\vec{u}'=0
}
\end{array}
\right .
\label{eqsdbis}
\end{equation}
and $\vu'=\vzero$ at $r=1$.

When $E=0$, these equations are solved by

\beq \vu' =  \lp -\frac{u_e}{r^2}+2\csd P_2(\cth)\rp\er+\csd r \dntheta{P_2}\etheta
\eeqn{mer_circ}
where $P_2(\cth)$ is the order 2 Legendre polynomial.  This flow
does not meet the inviscid boundary conditions $\vu'\cdot\vn=0$ at
$r=1$, however, it may be viewed as the meridional circulation associated
with a differential rotation.  In this case we need to evaluate the
pumping\footnote{Usually, there is a mass flux between a boundary
layer and its environment. This mass flux called ``pumping'' may be in
both direction. It comes from the fact that the horizontal variations
(horizontal here means parallel to the boundary) of the horizontal
components of the velocity may not verfiy mass conservation. Thus, a
small velocity of order of the non-dimensional thickness of the layer,
orthogonal to the layer, must be added.} of the Ekman layer. In the
boundary layer, the flow is $u_0+\tilde{u}_0$, where the boundary layer
correction $\tilde{u}_0$ is given by

\begin{equation}
(\vn\wedge\tilde{u}_0+i\tilde{u}_0)=-(\vec{n}\wedge
u_0+iu_0)_{\alpha=0}\exp(-(1+i)\alpha)
\end{equation}
where we dropped the primes (in $u$) and where

\[ \displaystyle{\alpha=\zeta\sqrt{\frac{|\cos
\theta|}{2}}=(1-r)\sqrt{\frac{|\cos \theta|}{2E}}}\]
\cite[e.g.][]{green69}. Identifying the $\theta$ and $\varphi$ components
of the velocity we get:

\beq
\left \{
\begin{array}{ccl}
\displaystyle{u_\theta}&=&\displaystyle{-U(\sin  \theta)\ \sin  \alpha\
e^{-\alpha}}\\
\displaystyle{u_\phi}&=&\displaystyle{U(s)-U(\sin\theta)\  \cos  \alpha\
e^{-\alpha}}
\end{array}
\right .
\eeqn{bls}
where the function $U(s)$ is the azimuthal (zero-order) component of
the geostrophic flow, which just depends on $s$, the radial cylindrical
coordinate, as imposed by Taylor-Proudman theorem\footnote{Taylor-Proudman
theorem states that when viscosity is negligible and no forcing applies,
the vorticity equation \eq{eqsd} leads to $\na\wedge(\ez\wedge\vu)=\vzero$
or $\partial_z\vu=\vzero$, meaning that the flow does not depend on the
coordinate along the rotation axis.}. The differential equation verified
by $U(s)$ is derived from mass conservation in the boundary layer.
Indeed, we know that

\beq -u_e+2\csd P_2(\cth)+\tilde{u}_r = 0 \at r=1\eeqn{blp}
A $\zeta$-integration of the continuity equation leads to

\[ \tilde{u}_r(\zeta=0) = \frac{1}{\sin
\theta}\sqrt{\frac{E}{2}}\frac{\partial}{\partial \theta} \left (\frac{\sin
\theta\ U(\sth)}{\sqrt{|\cth|}} \rp
\]
Using \eq{blp}, we finally get 

\beq
U(s)= -\csd\sqrt{\frac{2}{E}}\frac{(1-s^2)^{3/4}}{s}\lp\frac{u_e}{\csd} + s^2\rp
\eeqn{rotdiffsd}
From solution \eq{rotdiffsd} and expressions \eq{bls}, the viscous
torque applied to the outer turbulent layer can be evaluated. It gives
the nondimensional constant $k$ of \eq{defk}. It turns out that 

\[ k= \frac{8\pi}{15}\lp 1-\frac{1}{\beta}\rp\]
and

\beq u_e=-\frac{2\csd}{5\beta}\eeqn{ue}
therefore

\beq  U(s) = -\csd\sqrt{\frac{2}{E}}
\frac{(1-s^2)^{3/4}}{s}\lp-\frac{2}{5\beta} + s^2\rp \; .
\eeqn{rotdiffsdb}
We note that this solution is singular on the rotation axis because of the
singular nature of the outflow at $r=0$. In the numerics, we remove this
singularity by assuming that the outflow starts at some finite radius
(a core-envelope boundary here at $r=0.15$). Fig.~\ref{u_geo} shows a
comparison between the analytic and numerical solutions. At $E=10^{-6}$,
the difference in the envelope is hardly perceptible.  Fig.~\ref{u_merid}
illustrates the associated meridional circulation.

\begin{figure*}
  \begin{center}
\centerline{
\includegraphics[scale=0.3]{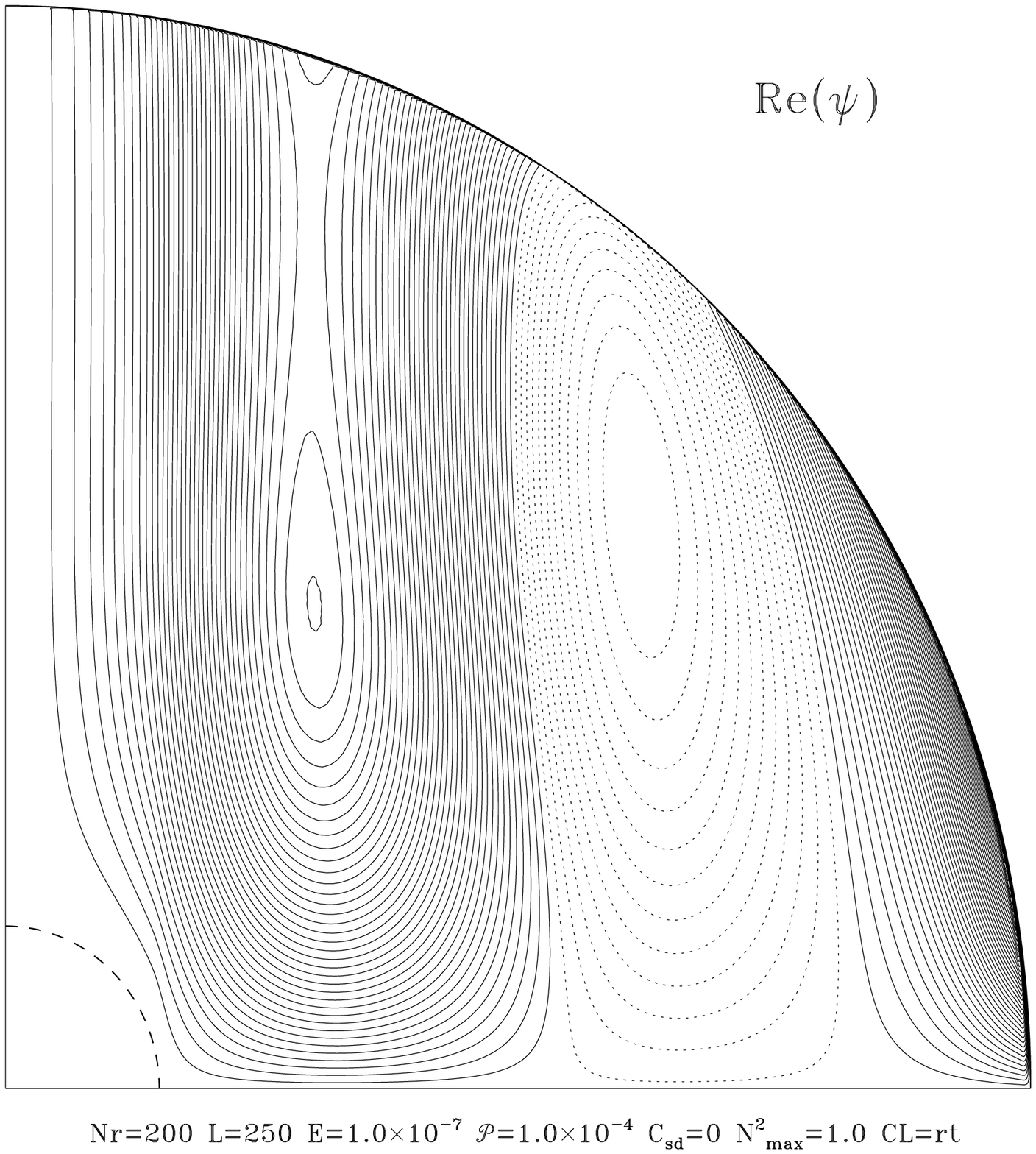}
\includegraphics[scale=0.3]{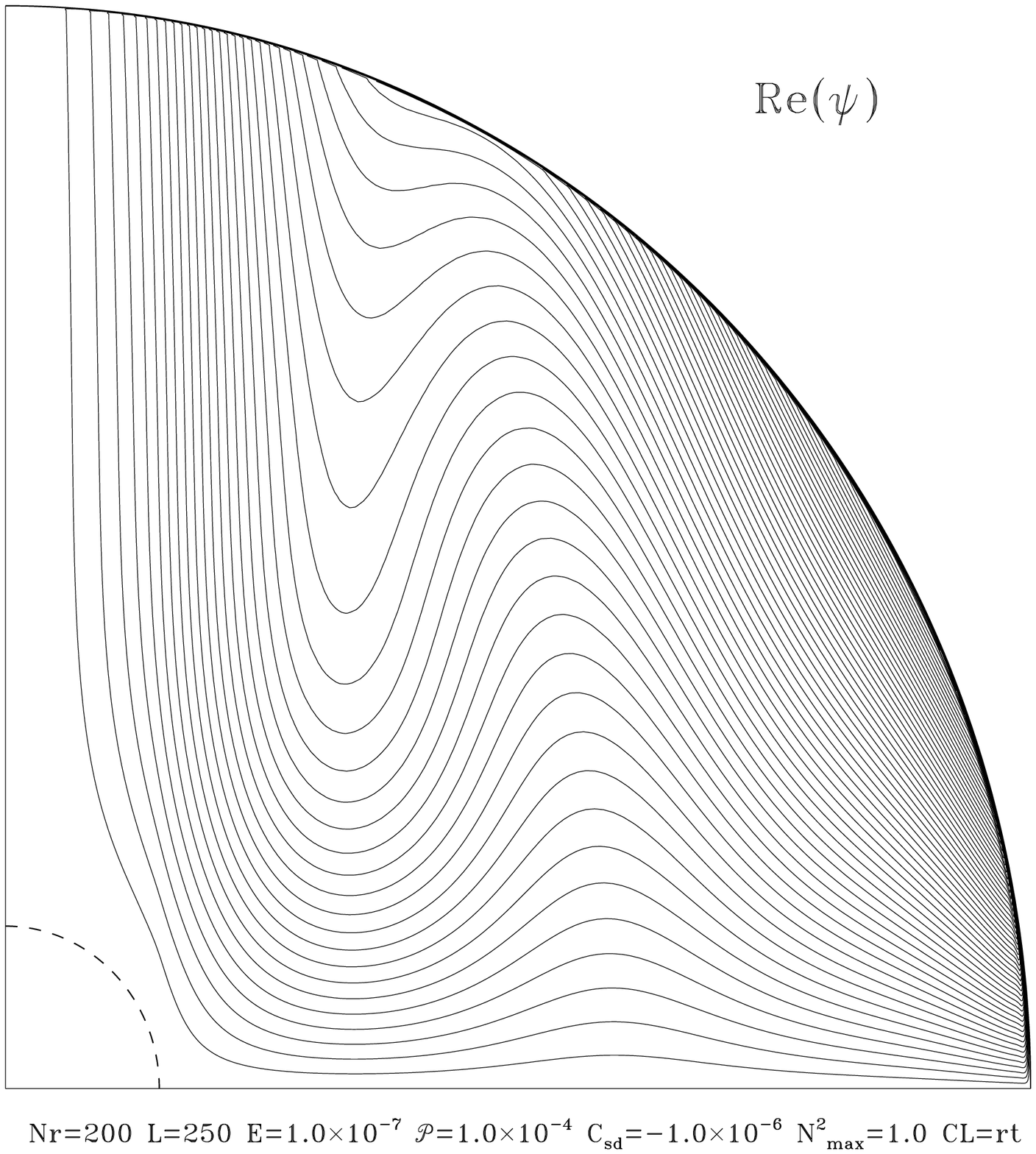}
\includegraphics[scale=0.3]{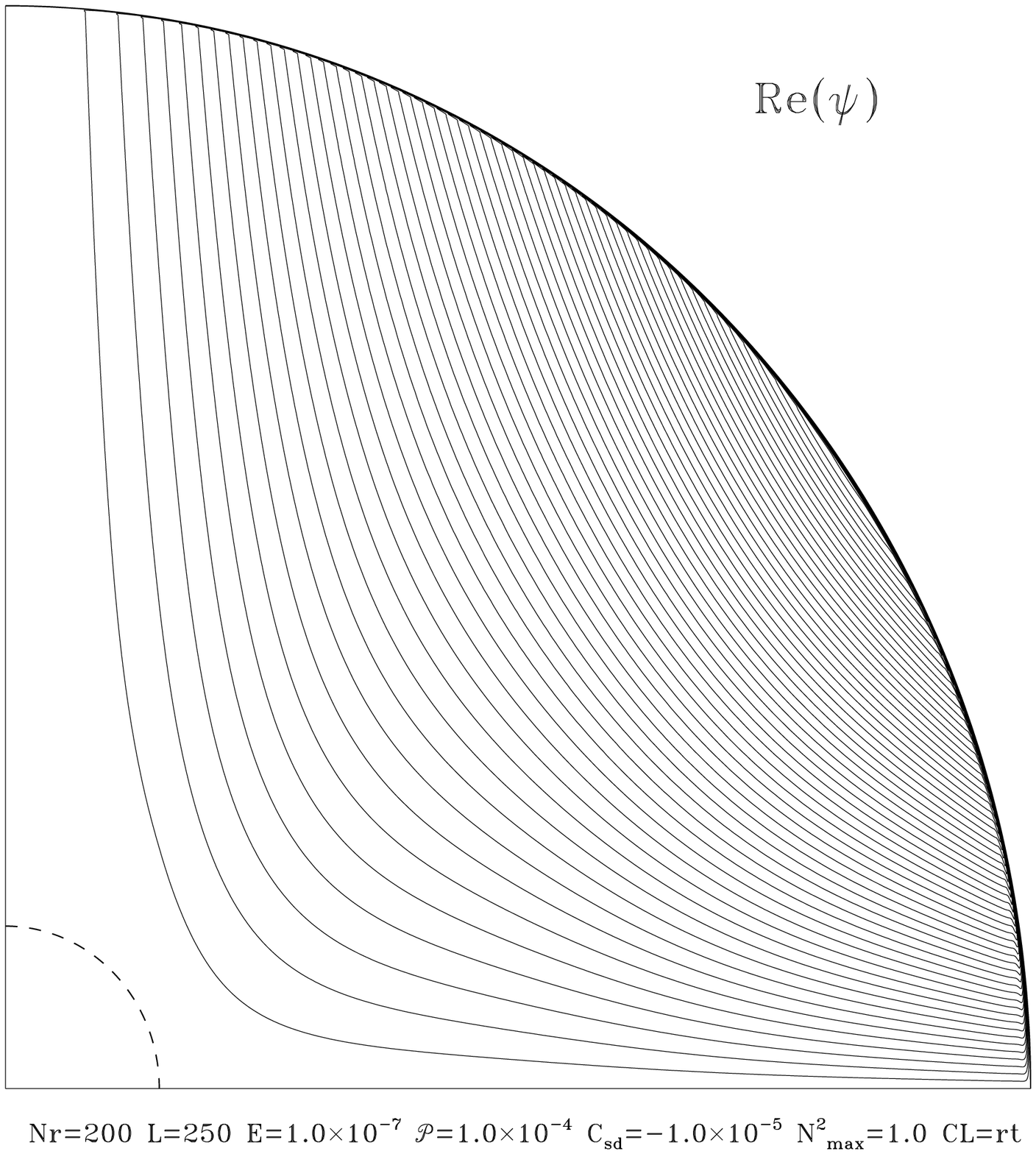}
}
  \end{center}
\caption{
Meridional circulation in  a stably stratified rotating fluid in a
sphere when the spin-down forcing $\csd$ is increased ($\csd = 0,
10^{-6}, 10^{-5}$ from left to right). Streamlines of the meridional
circulations are shown in the same way as in Fig.~\ref{u_merid}.  As in
Fig.~\ref{transit}, $E=10^{-7}$, $Pr=10^{-4}$, and $\beta\gg1$.
Numerical resolution is the same as in Fig.~\ref{transit}.}
\label{transitpsi}
\end{figure*}

The foregoing analytical solution does not take the effects
of buoyancy into account. However, these effects can be neglected in some range
of parameters. Let us first observe that the meridional circulation
induced by the spin-down should be associated with a temperature
fluctuation that verifies

\[ \theta_T \sim \frac{\csd}{\tilde{E}_T} = \csd\lambda/E \]
as given by the energy equation in \eq{eqstage}.
Since the Coriolis term is \od{\csd E^{-1/2}}, the buoyancy can be
neglected in the momentum equation if $\theta_T\ll\csd E^{-1/2}$ or when

\beq \lambda\ll\sqrt{E}\; . \eeqn{bound_sd}
This constraint is obviously not met in radiative region of the 3\msun\ 
star, but can be met in massive stars rotating near breakup as shown
by the 7\msun\ model numbers.

We now further explore the
properties of solution \eq{rotdiffsdb} in order to have a reference for
numerical solutions. It will also turn out that some properties carry
on in the domain of full coupling when $\lambda\supapp\sqrt{E}$.

\subsubsection{Baroclinicity versus spin-down}

The foregoing results allow us to determine the range of parameters
where either baroclinicity or spin-down dominate the driving of
the flows.

With the scaling leading to \eq{eqstage}, the differential rotation
arising from baroclinicity is of order unity \cite[see][]{R06}. Thus,
from \eq{rotdiffsdb}, we see that the differential rotation driven by the
spin-down dominates if

\beq \frac{\csd}{\sqrt{E}} \gg 1 \; .\eeqn{fineq}
Joining this condition with \eq{bound_sd}, we find that the influence
of a stable stratification on the differential rotation triggered by a
spin-down is negligible when

\beq \csd \gg \sqrt{E} \gg \lambda\; .\eeq
Besides, the meridional circulation associated with the spin-down flow
is \od{\csd} as given by \eq{mer_circ}. It overwhelms
the baroclinic circulation, which is \od{E}, if

\beq \csd \gg E\eeqn{seneq}
Since $E\ll1$, when \eq{fineq} is met, \eq{seneq} is also met. In this
case, the flow is completely dominated by the spin-down flow. For a
weaker spin-down, such that

\beq E\ll \csd \ll \sqrt{E}\eeqn{weakw}
the meridional circulation is triggered by the spin-down, while the
differential rotation is essentially coming from the baroclinic torque.
This result shows that the meridional circulation driven by baroclinicity
is actually extremely weak because Ekman numbers are usually less than
10$^{-8}$. 

The foregoing inequalities show that three regimes may be distinguished:
a {\em strong wind regime} occurs when \eq{fineq} is verified, namely when
the spin-down flows dominate both the circulations and the differential
rotation, a {\em moderate wind regime} described by \eq{weakw},
when meridional circulation is that imposed by the spin-down and the
differential rotation is controlled by baroclinicity, and finally a weak
or null-wind regime when baroclinic flows are only slightly perturbed
by the spin-down.

In Figs.~\ref{transit} and \ref{transitpsi}, we illustrate this case where
$\lambda\ll\sqrt{E}$ so that the buoyancy is negligible and the analytic
solution applies. We choose $\lambda=10^{-4}$ and $E=10^{-7}$ for various
values of $\csd$.  The figures clearly illustrate the transitions that are
expected from analytics, namely that, as the spin-down forcing increases,
the meridional circulation first transits to the spin-down meridional
circulation around
$\csd=10^{-6}$, while the differential rotation reaches the asymptotic
state enforced by spin-down when $\csd\supapp\sqrt{E}$.

\begin{figure*}
  \begin{center}
\centerline{
\includegraphics[scale=0.3]{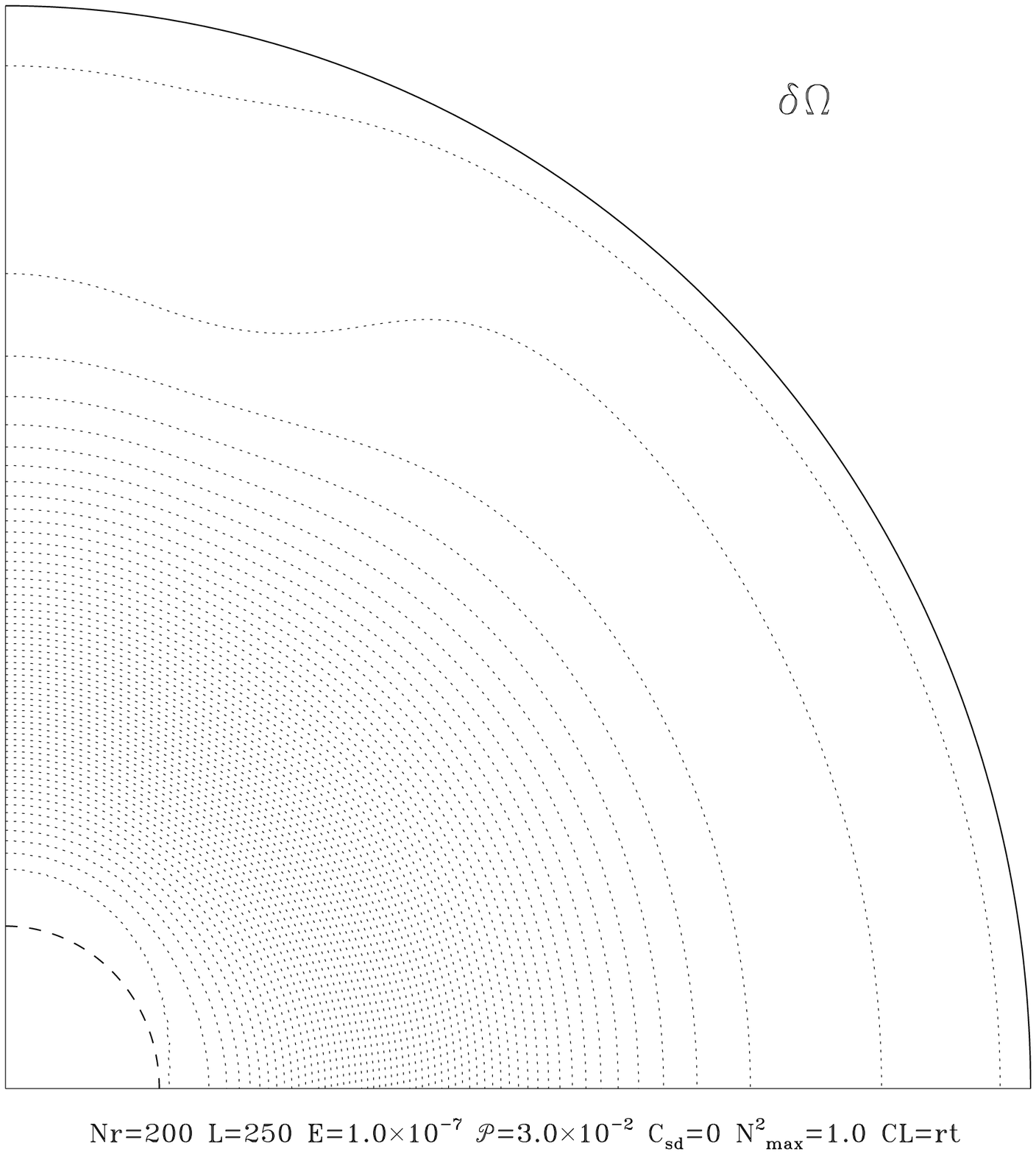}
\includegraphics[scale=0.3]{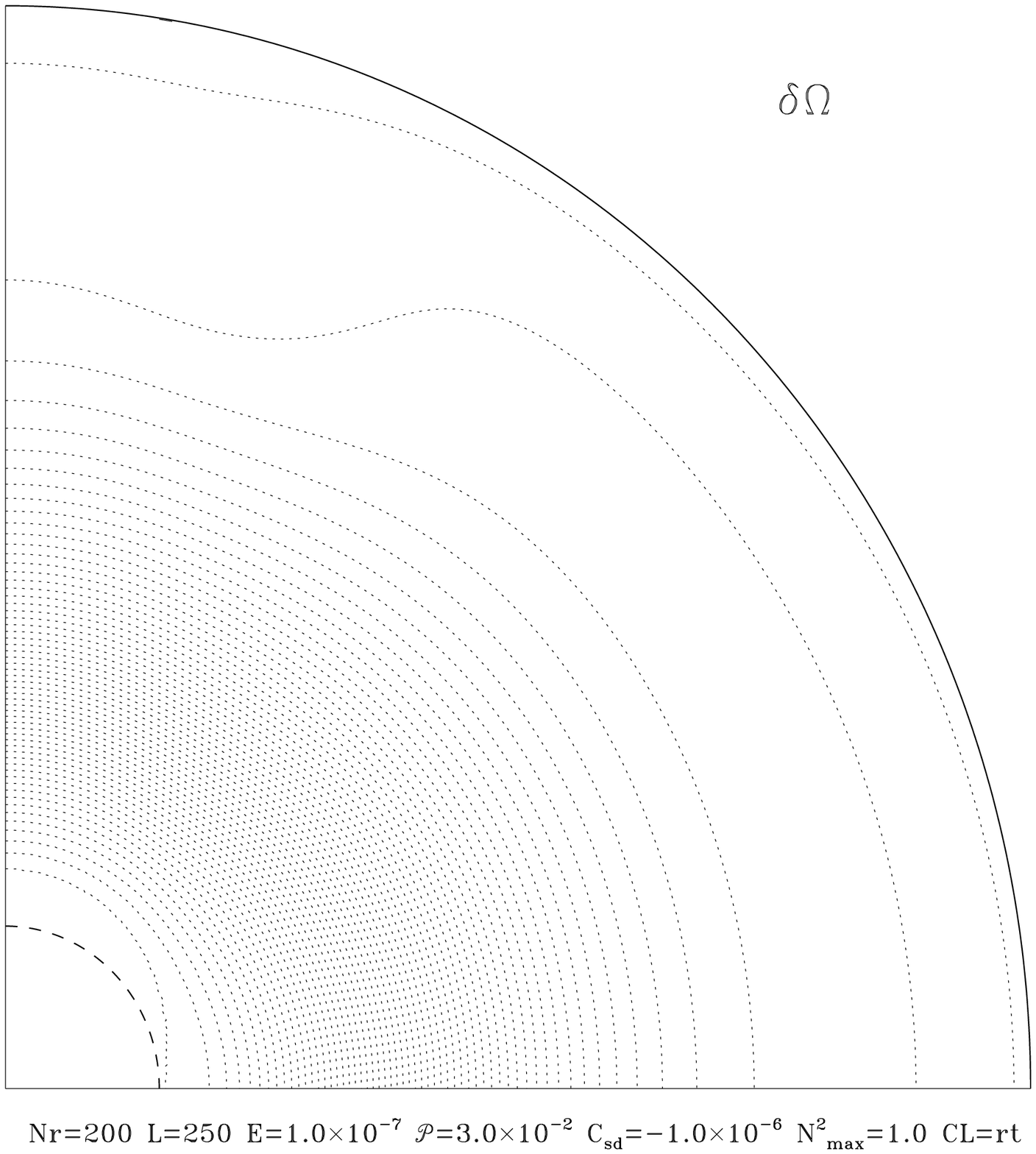}
\includegraphics[scale=0.3]{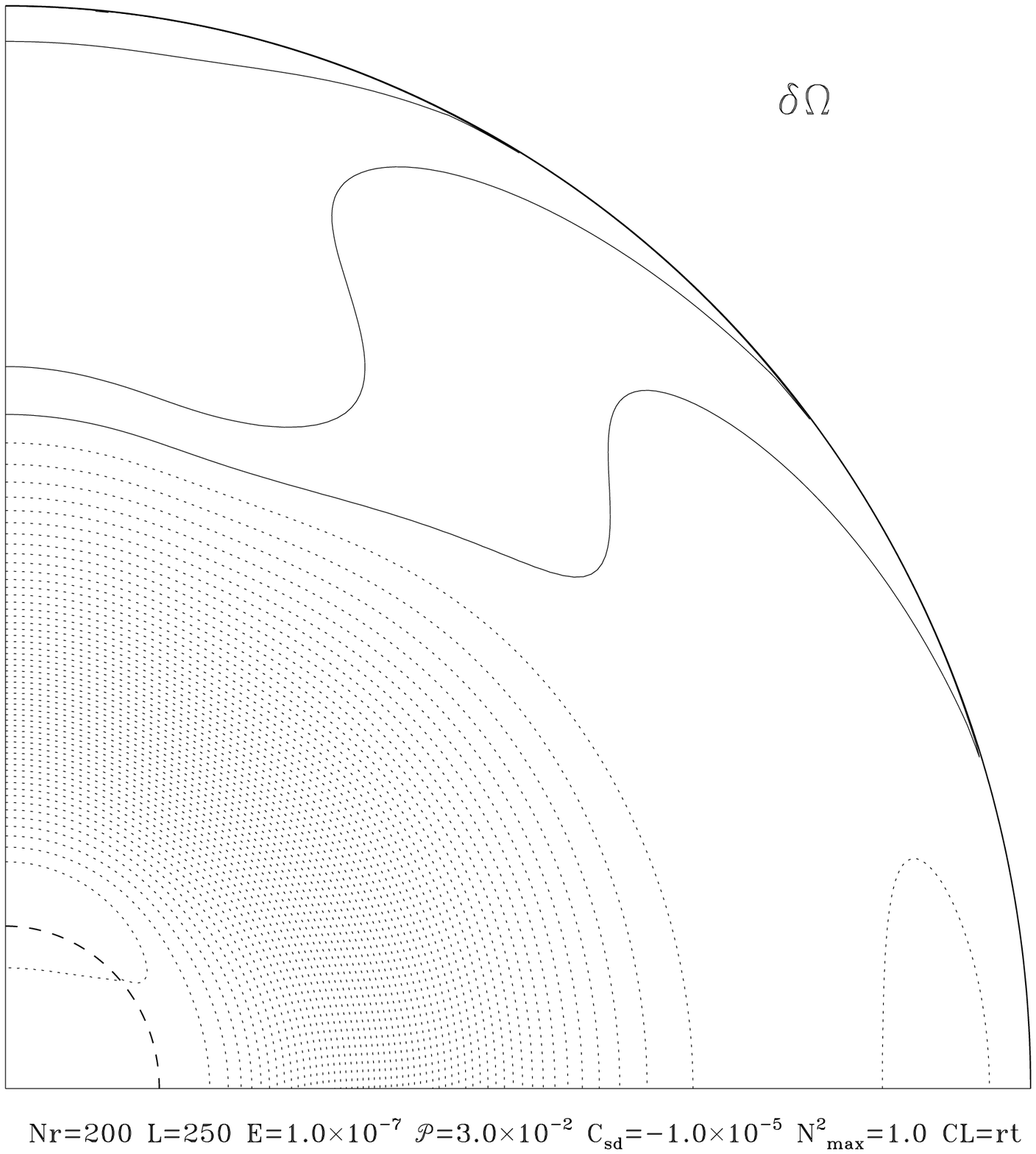}
}
\centerline{
\includegraphics[scale=0.3]{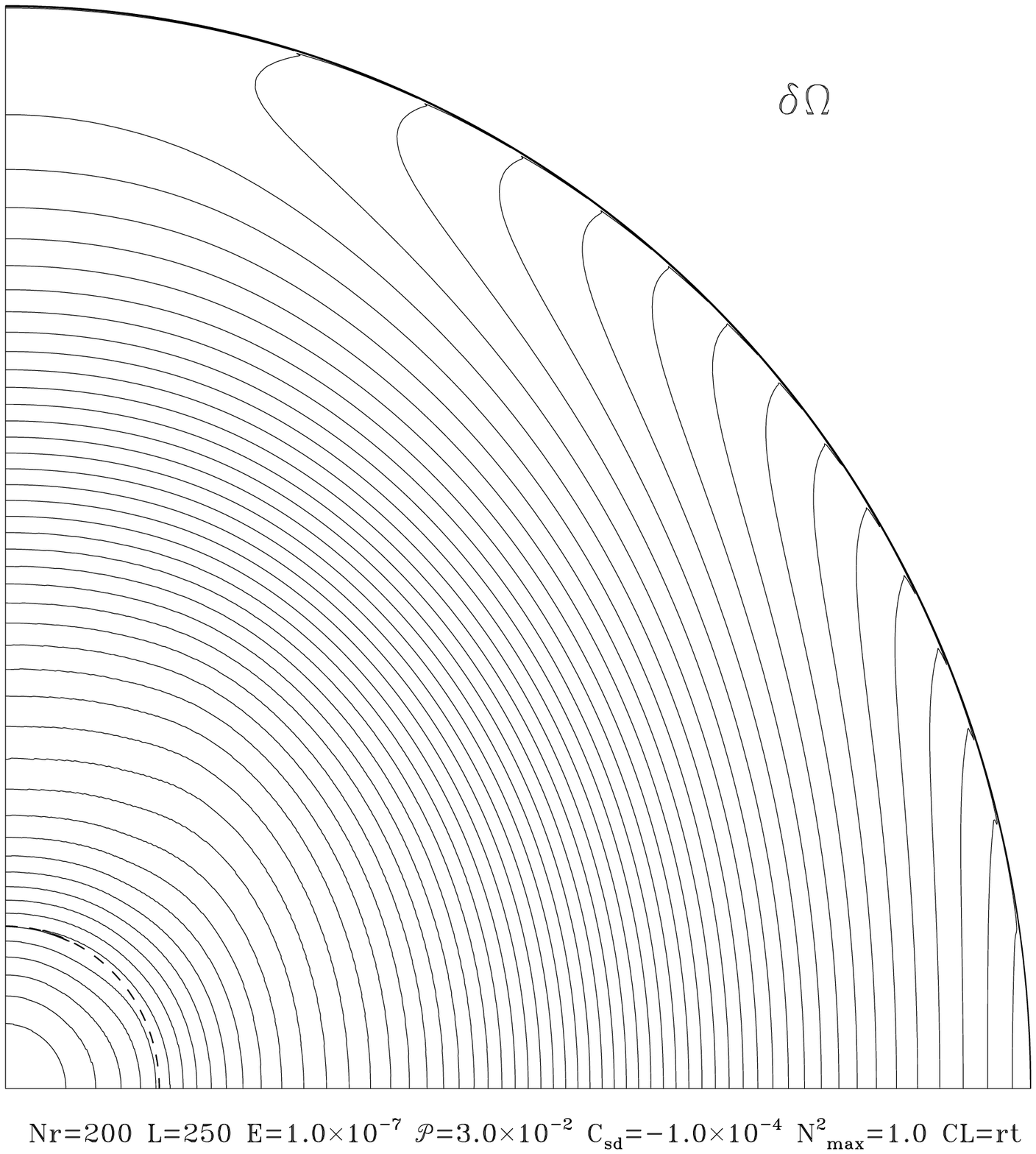}
\includegraphics[scale=0.3]{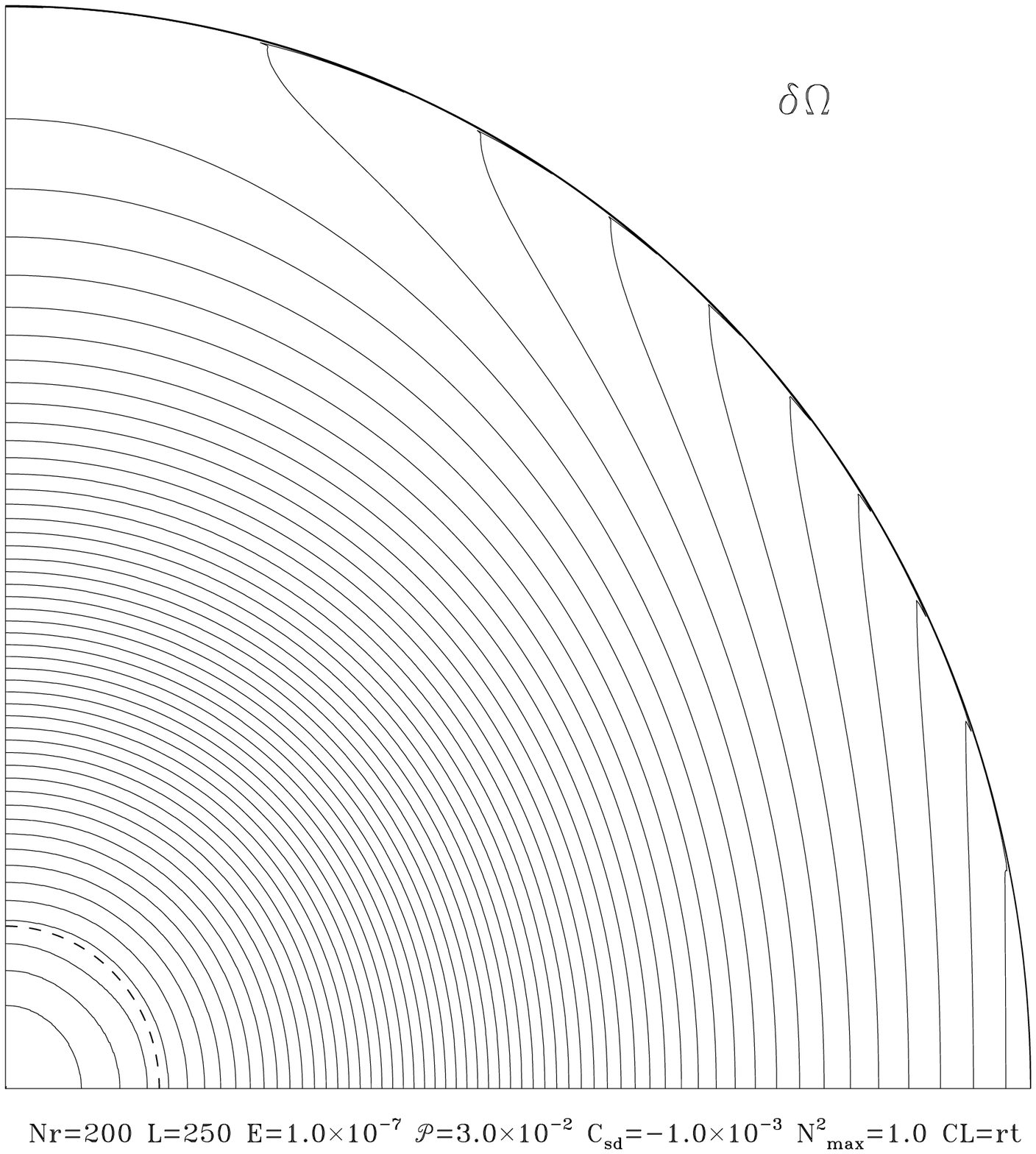}
\includegraphics[scale=0.3]{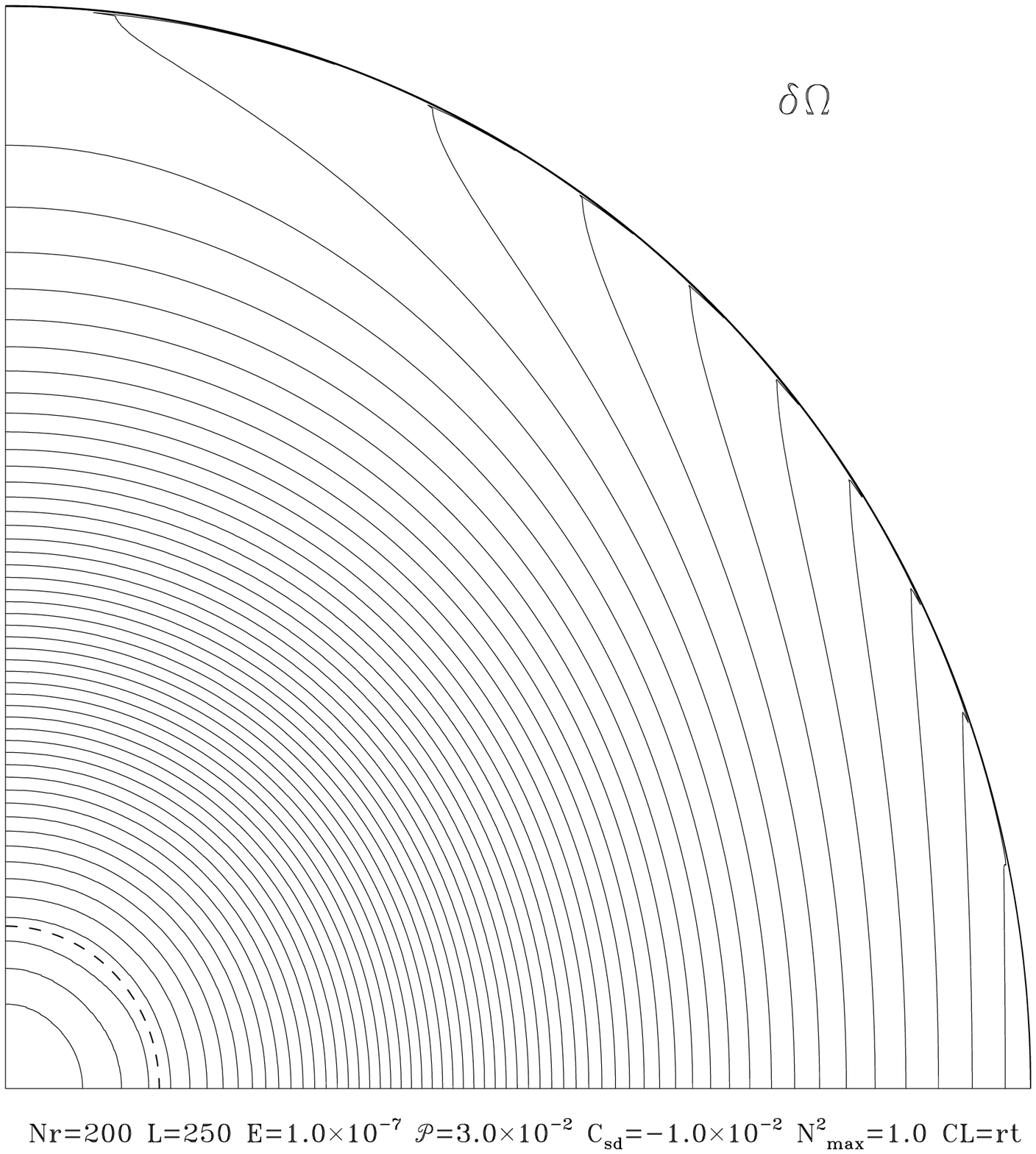}
}
  \end{center}
\caption{
Differential rotation when the spin-down forcing $\csd$
is strengthened while the buoyancy is influential. From left to right and top to bottom $\csd=0,
-10^{-6}, -10^{-5}, -10^{-4}, -10^{-3 }, -10^{-2}$
We use $E=10^{-7}$, $Pr=3\times10^{-2}$,
$\lambda\simeq3\times10^{-2}=100\sqrt{E}$ and $\beta\gg1$. Top row:
the flow is dominated by the baroclinic forcing.  Bottom row: the flow
is dominated by the spin-down forcing. Numerical resolution is the
same as in Fig.~\ref{transit}.}
\label{transit100}
\end{figure*}

\subsubsection{Centrifugal instability}\label{centr_inst}

With the analytic expression of the azimuthal velocity \eq{rotdiffsdb}
we can determine the conditions of the appearance of the centrifugal
instability. Indeed, when the specific angular momentum of the fluid
decreases with the distance to the axis, axisymmetric disturbances
can grow. Noting that the specific angular momentum $\ell$ reads

\[ \ell = s^2+sU(s),\]
the condition $d\ell/ds\leq 0$ leads to

\beq (1-s^2)^{1/4} \leq
|\csd|\sqrt{\frac{2}{E}}\lp\frac{7}{4}s^2-1 +\frac{3}{10\beta}\rp \; ,
\eeqn{cfi}
which determines the region where the centrifugal instability develops.

If $|\csd|\ll\sqrt{E}$, namely in the moderate wind condition, \eq{cfi} only
applies to a very small fraction of the volume. Indeed, the centrifugal
instability develops when

\[s\geq1-\od{\csd^4/E^2}\; .\]
If we observe that this condition applies outside the Ekman layer, we
need demanding $\csd^4/E^2>E^{1/2}$ or that

\[ E^{5/8} < \csd < E^{1/2}\; ,\]
which quite restricts the range of acceptable $\csd$-values. Hence, in
the moderate wind regime, the influence of the centrifugal instability is
likely marginal.

On the other hand, in the strong wind regime, we only demand that the
rhs of \eq{cfi} be positive so that all the volume beyond

\[ s_m = \frac{2}{\sqrt{7}}\lp1-\frac{3}{20\beta}\rp\]
can develop the centrifugal instability. This region of the star is
sketched out in Fig.~\ref{croquis} for infinite $\beta$. We thus expect
that equatorial regions are more mixed than the polar regions.

\begin{figure*}
  \begin{center}
\centerline{
\includegraphics[scale=0.3]{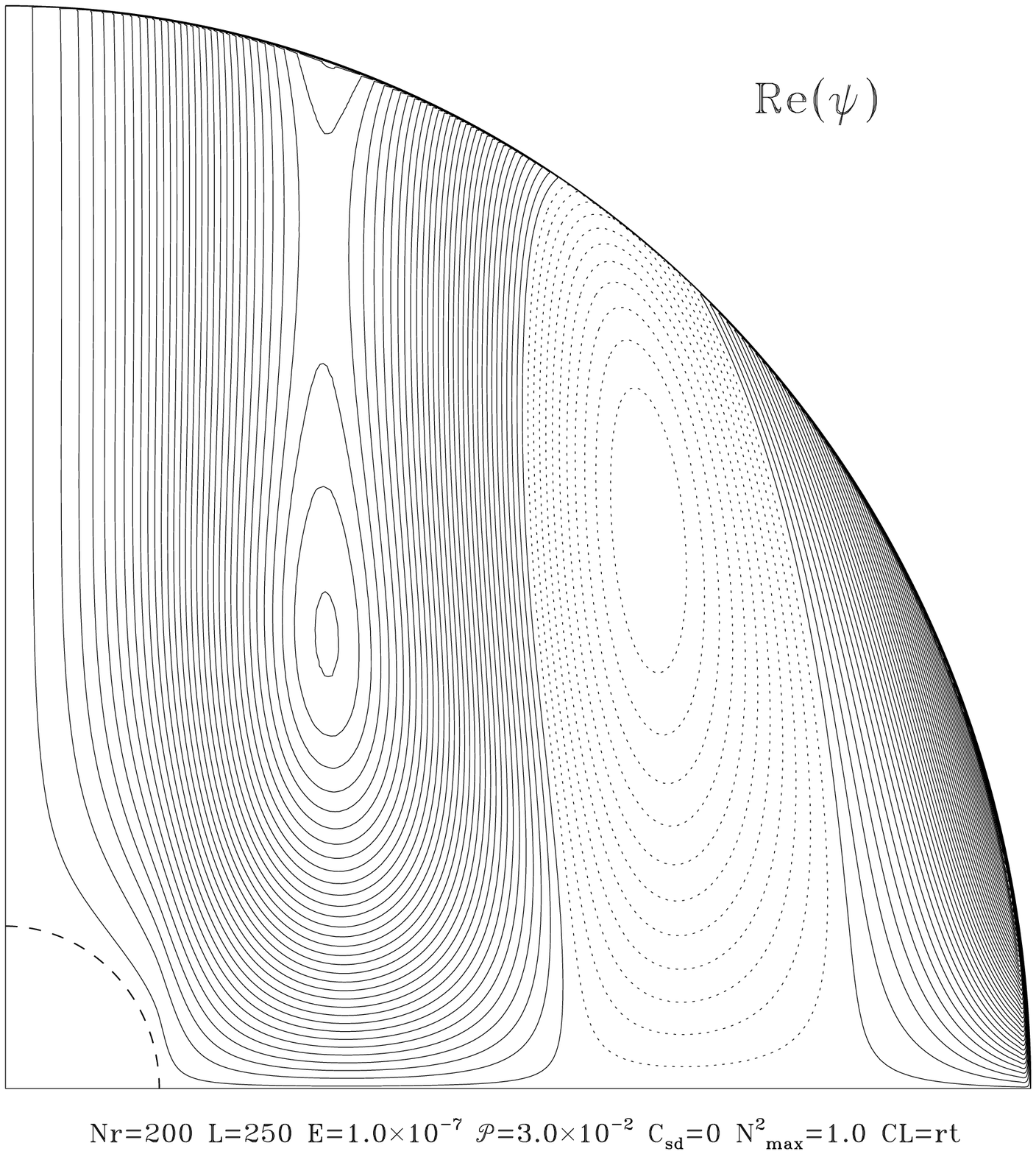}
\includegraphics[scale=0.3]{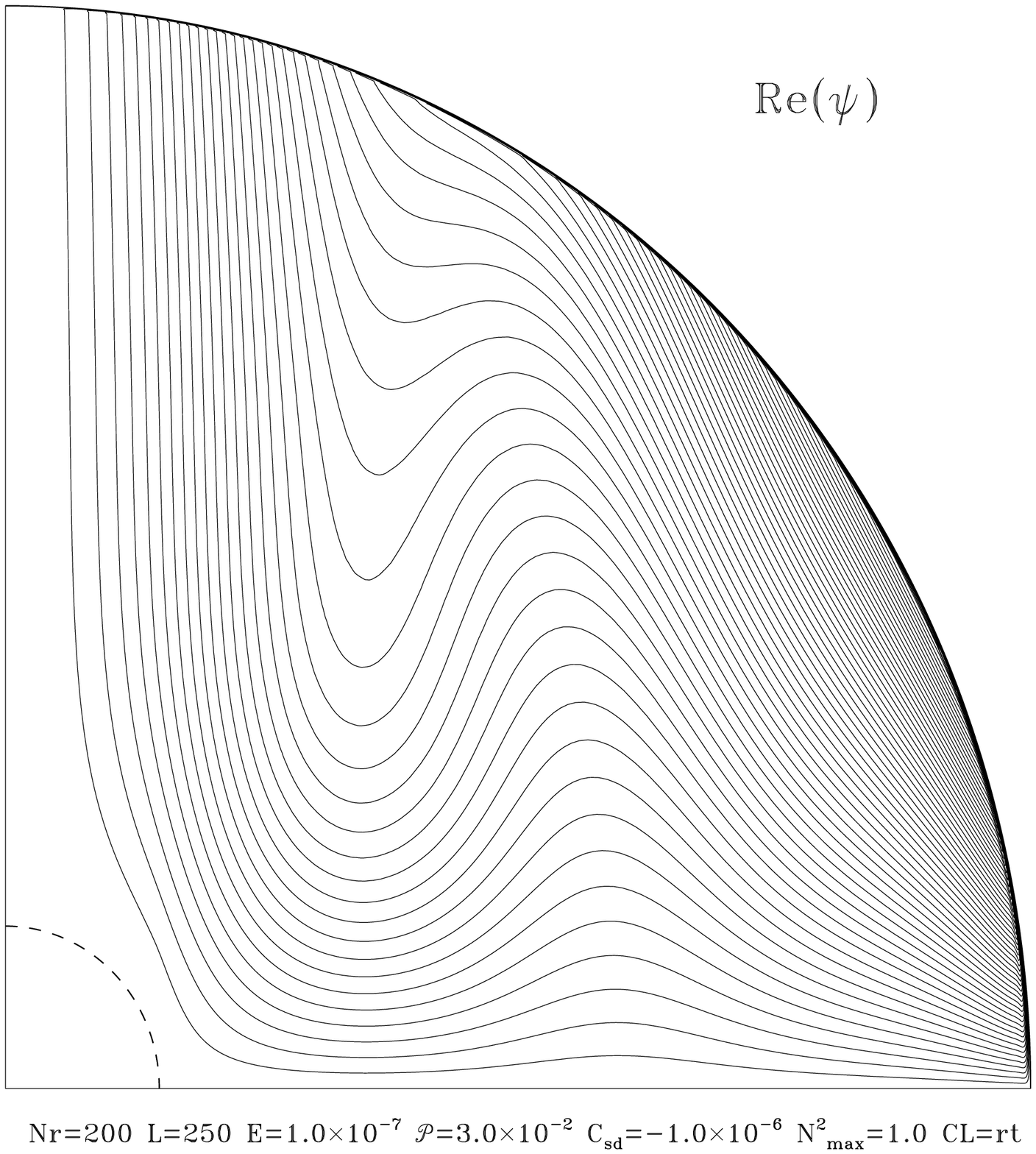}
\includegraphics[scale=0.3]{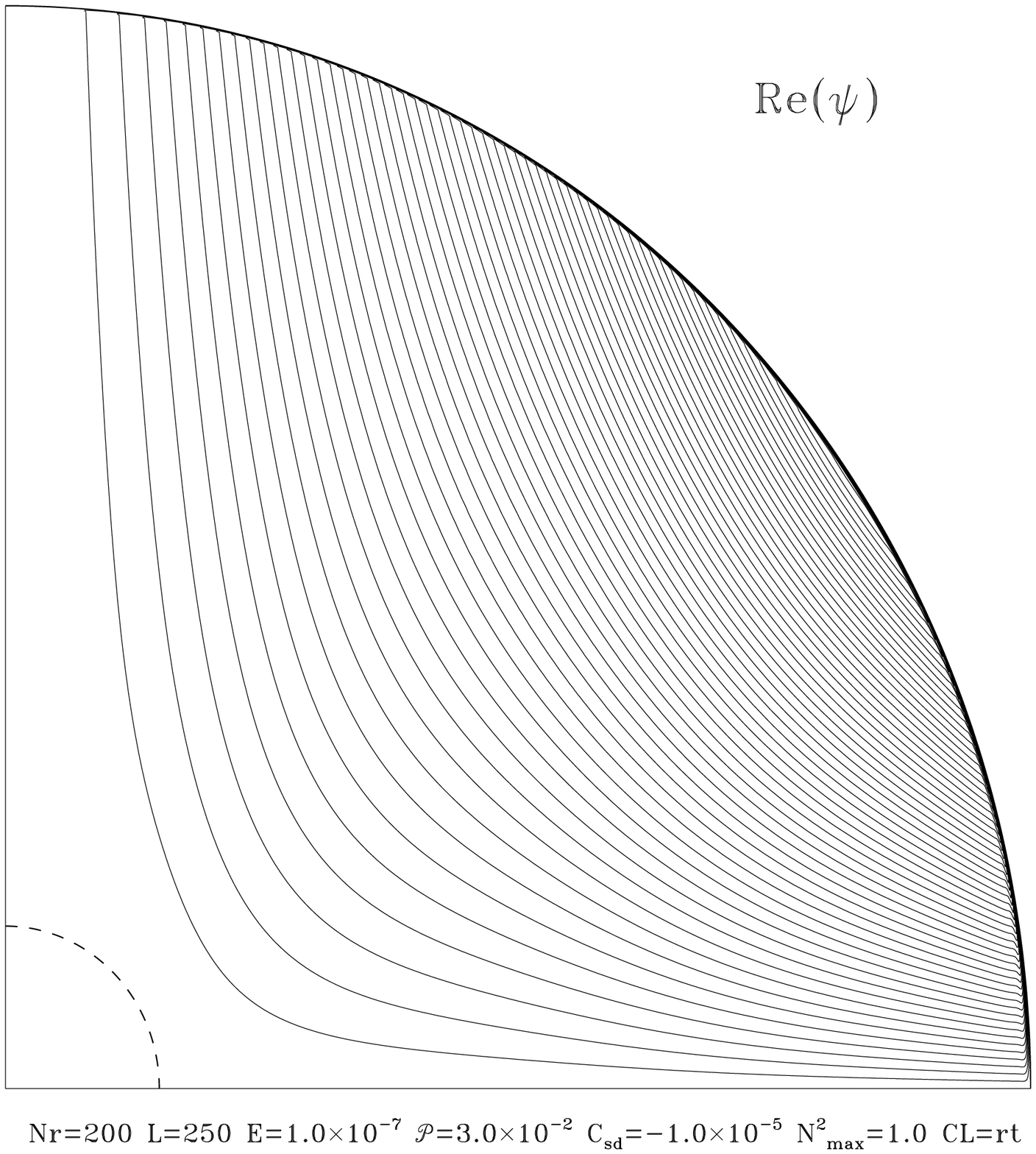}
}
  \end{center}
\caption{
Change in the meridional circulation when the spin-down forcing is
increased. We represent the streamlines of the meridional circulation
for $\csd=0, 10^{-6}, 10^{-5}$ from left
to right.  As in Fig.~\ref{transit100} $E=10^{-7}$, $Pr=3\times10^{-2}$
and $\beta\gg1$. Solid line show counter-clockwise circulation.
Numerical resolution is the same as in Fig.~\ref{transit}.}
\label{transitpsi100}
\end{figure*}

\subsubsection{The influence of buoyancy}

The foregoing derivation neglected the coupling of the spin-down flow
with the temperature field, which occurs through buoyancy. As already
noticed, this is not possible in the radiative region of stars
when $\lambda\gg\sqrt{E}$.

To investigate the fully coupled case, we resort to numerical solutions
of the complete system \eq{eqstage}.  The numerical method is the same
as in \cite{R06} and will not be repeated here. It is
based on a spectral method that uses spherical harmonics to represent
the angular variations of the solutions and Chebyshev polynomials for
the radial dependence.

In Fig.~\ref{transit100}, as in Fig.~\ref{transit}, we investigate
the transition from a pure baroclinic flow to a spin-down dominated
flow. Obviously, the flow also makes this transition for values
$\csd\sim\sqrt{E}$ but reaches a new state close to a shellular rotation,
especially in the central region.

The meridian streamlines, depicted in Fig.~\ref{transitpsi100}, show a
transition to the spin-down dominated state at a lower value of $\csd$,
as in the uncoupled case.

The foregoing shellular rotation can be understood from \eq{eqstage},
if we neglect the viscosity. The vorticity equation now reads:

\[ \dz{\vu} = (\partial_\theta\vartheta+n^{2}(r)\sth\cth)\ephi +2\csd\ez
\; .\]
Considering that the meridional circulation \eq{mer_circ} compensates the
spin-down torque $2\csd\ez$, we note that the radial component reads
$u_r=\csd r (3\cos^2\theta-1) = 2r\csd P_2(\cos\theta)$.
We easily see that the
temperature perturbation associated with this circulation reads:

\[ \vartheta = \frac{\csd}{\tE_T}\vartheta_2(r)P_2(\cos\theta)\]
where $\vartheta_2(r)$ is a \od{1} function determined by the \BVF\
profile $n^2(r)$.  When the amplitude of the temperature perturbation
$\csd/\tE_T$ is larger than unity, the baroclinic torque generated by
the spin-down flow overwhelms the baroclinic torque resulting from the
centrifugal force, $-n^{2}(r)\sth\cth\ephi$.  This leads to an
azimuthal velocity that may be written

\[ u_\varphi = \frac{\csd}{\tE_T}\Omega(r)\sth+U(s)\]
where $U(s)$ is the geostrophic flow that makes the boundary conditions
verified and $\Omega(r)$ is determined by $\vartheta_2(r)$. If
$U$ is small enough, we see that a shellular differential rotation
naturally emerges. This seems to be the case for the values chosen in
Fig.~\ref{transit100}.

If $u_\varphi$ is \od{\csd/\tE_T}, internal balance of
viscous force and Coriolis force gives a radial flow that is
\od{E\csd/\tE_T} or \od{\lambda\csd}. On the other hand, Ekman pumping
generates a \od{\sqrt{E}\csd/\tE_T} or \od{\lambda\csd/\sqrt{E}}
circulation. Numerical solutions show that the baroclinic circulation
still disappears when $\csd>E$, indicating that Ekman pumping is weak (it
can be zero if the latitudinal flow varies appropriately) and therefore
that condition \eq{seneq}, determining the domination of the spin-down
driven circulation, likely extends for all $\lambda$ less than unity.

\subsection{Stress-driven spin-down}

The case of a stress-driven spin-down has been fully analysed by
\cite{fried76}, focusing on the slowing down rotation of the
radiative zone of the Sun. We shall not repeat this complex analysis,
but focus directly on our original question as to which condition
characterizes the dominance of spin-down circulation compared to
the baroclinic circulation.

For this, we reconsider the boundary layer analysis of \cite{R06}
in sect.~3.2. The stress-free boundary conditions are now modified into

\beq \dr{}\lp\frac{u_\theta+iu_\varphi}{r}\rp= -i\tau(\theta)\eeqn{bc2}
where $\tau$ is the non-dimensional surface stress. The general expression
of the flow in the Ekman layer is

\[ u_\theta+iu_\varphi = C\exp\lp-\zeta\sqrt{i|\cth|}\rp +
iu_\varphi^0\]
where $u_\varphi^0$ is the interior inviscid solution that reads

\[ u_\varphi^0 = s\int \frac{n^2(r)}{r}dr + F(s) \; .\]
The constant $C$ is such that condition \eq{bc2} is met. This leads to
\[ C\equiv C(\theta)= (1+i)\sqrt{\frac{E}{2}}\Gamma(\theta)\]
with

\begin{equation}
\Gamma(\theta) = \frac{F(\sth)-\sth
F'(\sth)-n^2(1)\sth -\tau(\theta)}{\sqrt{|\cth|}}
\end{equation}
where the prime indicates a derivative.

From this latter expression, it is clear that the stress driving will
overtake the baroclinic driving when

\[ \tau \gg n^2(1) \sim 1\; .\]
Indeed, the scaling has been chosen such that $n^2(r)$ is \od{1}.
We further note that when this inequality is met,  both the circulation
and the differential rotation overtake their baroclinic equivalent.
This is because both flows meet boundary conditions on the stress.

As shown by \cite{fried76}, the radial driving of the circulation
by the pumping of the boundary layer is similar as in the velocity
driven case if $\tau(\theta)=\tau_e\sth$. In this case, one finds

\[ \tu_r = -E\tau_eP_2(\cth)\]
similar to expression \eq{blp} when $\beta\gg1$.

\section{Discussion}

We now replace the foregoing results in the astrophysical context.

\subsection{Stress-driven spin-down}
\subsubsection{A transition mass-loss rate}

We may estimate the stress imposed by the turbulent layer if we
follow the result of \cite{LCM00} that the angular velocity profile is
such that the specific angular momentum in the layer remains constant. In
such a case,

\beq \Omega_l(s)\propto s^{-2}\; ,\eeqn{omprof}
and the azimuthal stress is 

\[ \tau_* = \mu_ts\ds{}\frac{v_\varphi}{s}=-2\mu_t\Omega_l(s)\]
so that at the interface

\[ \tau_* = -2\Omega\mu_t\]
where $\Omega$ is the angular velocity of the fluid at the interface
and $\mu_t$ is the turbulent viscosity.

If we further assume that the turbulent layer propagates
in a stably stratified envelope without removing the stable
stratification\footnote{This means that the P\'eclet number Pe of
this turbulence is small compared to unity. In other words, turbulent
diffusion remains small compared to radiative diffusion. With Zahn's
model, Pe=$v\ell/\kappa\sim \nu_T/\kappa\sim(2\Omega/\calN)^2/12$.  Using
stellar data of Tab.~\ref{param_stars}, we find Pe$\infapp4\times10^{-3}$
in all cases. This is small indeed.}, we can use the turbulent viscosity
of Zahn's model \cite[][]{zahn92}, namely

\[ \mu_t = \rho\frac{\RI_c\kappa}{3}\lp\frac{s}{\calN}\dns{\Omega}\rp^2
\]
where $\RI_c$ is the critical Richardson number. Assuming that the angular
velocity profile verifies \eq{omprof} and that $\RI_c\sim1/4$, we get

\beq \mu_t\sim \frac{\rho\kappa}{12}\lp\frac{2\Omega}{\calN}\rp^2\; .
\eeqn{turbvisc}
Note that the place where the boundary conditions are taken is arbitrary
within the turbulent layer. The stress at the interface is

\beq
\tau_*\sim-\frac{\rho\kappa\Omega}{6}\lp\frac{2\Omega}{\calN}\rp^2 \; .
\eeqn{mut}
We note that the Ekman number associated with this turbulent viscosity
is

\beq E_t= \frac{\nu_t}{2\Omega R^2}  \sim \frac{\kappa}{24\Omega
R^2}\lp\frac{2\Omega}{\calN}\rp^2 = \tE_T/12\eeqn{ekt}
thus
\[ \lambda\sim 1/12\]
when turbulence is fully developed. In passing, we note that the ratio between
viscosity and turbulent viscosity is

\beq \frac{\mu}{\mu_t} \sim 12\lambda\; .\eeqn{mumut}
Turbulence  only increases momentum transport in fast rotating stars where
$12\lambda<1$ (typically rotation periods less than 30~days).

Expression \eq{mut} shows that the stress imposed by the turbulence is
determined by the local physical conditions of the interface.

We now reconsider the balance of angular momentum \eq{amb} together
with \eq{mut}. From the definition of $\beta$, we get the expression

\beq \beta= 1 + \frac{|\dotM_t|}{|\dotM|}\eeqn{beta}
where we introduced the ``transition'' mass-loss rate

\beq \dotM_t = \frac{2\pi}{3}\rho\kappa R\lp\frac{2\Omega}{\calN}\rp^2
\; .
\eeqn{Mtrans}
This expression shows that a characteristic mass flux exists
that determines whether  transport of angular momentum is dominated by
advection or turbulent viscous diffusion. This is a consequence of the
assumptions  \eq{omprof} together with the turbulence model. They make
the turbulent angular momentum flux due to diffusion depending only on
local conditions. Hence, the adjustment to the actual angular momentum
loss rate is made by advection. Our restriction $\beta\gg1$ therefore
selects mass losses smaller than $|\dotM_t|$. The range $|\dotM|>|\dotM_t|$
therefore naturally delineates a strong wind regime.

\begin{figure}
\begin{center}
\includegraphics[width=0.9\linewidth]{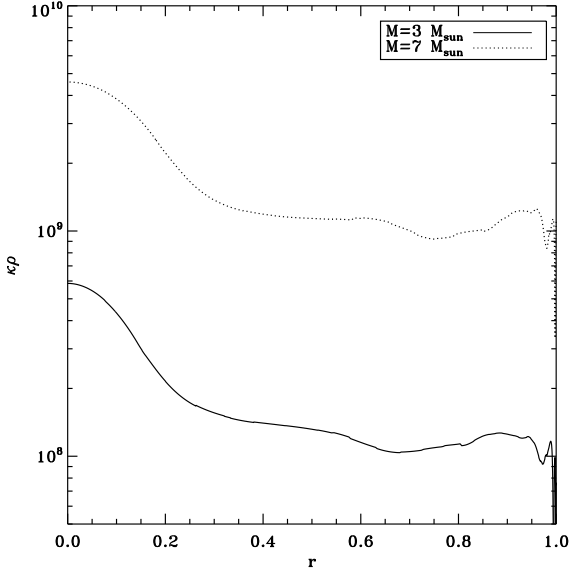}
\caption[]{The product of density with thermal diffusivity as a function
of the radius for our two stellar models.}
\label{kapparho}
\end{center}
\end{figure}

In Tab.~\ref{mopt}, we have computed the transition mass-loss rate for
our two stellar models.  Quite remarkably, we find that the values are
insensitive to the depth of the layer. This is because $\rho\kappa$
(the product of density and thermal diffusivity) is almost constant
in the upper half (in radius) of the star (see Fig.~\ref{kapparho}). We
find $\rho\kappa\sim 10^8$~cgs for the 3~\msun\ star and $\rho\kappa\sim
10^9$~cgs for the 7~\msun\ star.

\begin{table}
\begin{center}
\begin{spacing}{1.3}
\begin{tabular}{ccc}
\hline
Mass (M$_\odot$)& $\dot{M}_t(P_{\rm rot}=0.5$d) &$\dot{M}_t
(P_{\rm rot}=36$d)\\
3               & $10^{-8}$                    & $2\times10^{-12}$ \\
7               & $4\times10^{-7}$             & $7\times10^{-11}$ \\
\hline
\end{tabular}
\end{spacing}
\end{center}
\caption[]{Transition mass-loss rates (in M$_\odot$/yr) for the two models
and two rotation rates.}
\label{mopt}
\end{table}

The values of Tab.~\ref{mopt} show that for fast rotating stars rather
strong winds are necessary to make advection dominating.

\subsubsection{Baroclinic and spin-down flows}

We now assume that the bulk of the star is decelerated by the turbulent
stress of the upper layers, namely by

\beq \tau_*\sim2\mu_t\Omega \; .\eeqn{taustar}
The condition by which the spin-down flow supersedes the baroclinic
flows is that the non-dimensional stress (cf Eq.~\ref{taund}) is larger
than unity

\beq \tau_*\frac{2\Omega}{\mu\epsilon\calN^2}>1\eeqn{majtau}
or

\beq \nu_t> \nu\lp\frac{\calN}{2\Omega_k}\rp^2\eeqn{rapnu}
where $\Omega_k$ is the keplerian angular velocity at the layer's
radius.

If we use Zahn's prescription on the turbulent viscosity, we may
transform the previous inequality \eq{rapnu} into

\beq \lambda \infapp
\frac{1}{12}\lp\frac{2\Omega_k}{\calN}\rp^2\; .\eeqn{lamineq1}
For our two stars this inequality means 

\beq \lambda \infapp 0.02\eeqn{lamineq}
independent of their mass.

The foregoing inequality is not very stringent: in the case that we
are considering, only the slowly rotating models do not meet
this inequality. This threshold means that if turbulence, as described
by \eq{turbvisc}, exists, then the stress is large enough to remove
baroclinic flows.

When the mass-loss rate decreases, however, there must be a threshold
below which turbulence cannot be maintained. We surmise that this
occurs when the angular momentum flux that characterizes the baroclinic
flow equals the angular momentum loss of the star; hence

\[ \dotJ = \frac{2}{3}\dotM\Omega R^2\infapp 4\pi R^2V_r^{\rm baroc},\]
which also reads

\[ \dotM \infapp \dotM_c = \frac{3\pi}{2}\mu R
\lp\frac{\calN}{\Omega_k}\rp^2\; .\]
We may compare this critical mass-loss rate to the transition one and we find

\beq \dotM_c \sim 4\lambda\lp\frac{\calN}{\Omega_k}\rp^2\dotM_t \; .\eeqn{mdotc}
Since, for fast rotating models, $\lambda\sim 2\times10^{-5}$,
it turns out that $\dot{M}_c\sim 10^{-3}\dot{M}_{t}$ for these
stars, thus showing that a slight mass loss ($\sim 10^{-11}$\msun/yr)
may impose its dynamics on the stellar interior.

\subsection{Spin-down by a rigid layer}

We now turn to the other boundary condition where the spinning-down
layer imposes its velocity. This case might represent a turbulent layer
threaded by magnetic fields, which give some rigidity to the fluid.
In these conditions, we can consider the case of Sect.~\ref{rigcase}. To
further simplify the discussion, we assume that the layer is in a
turbulent state triggered by internal shear and that buoyancy
can be neglected so as to use the analytic solution \eq{rotdiffsd}. 

The novelty introduced by these boundary conditions is that the transition
from a baroclinic flow to the spin-down flow occurs in two steps.
When the mass loss is increased, the meridional circulation first changes
to that of the spin-down circulation. At a higher mass loss the differential
rotation of baroclinic origin leaves the place to that of spin-down
origin. The first threshold (meridional circulation) is reached when
$\csd>E$ according to \eq{seneq}. With the help of \eq{ue}, we find that
this condition is equivalent to

\beq \frac{5\beta|\dotM|\Omega}{4\pi R^3\rho\calN^2\eps}
>E\; .\eeqn{condrcap1}
In this expression $\beta$ is arbitrary. As shown by \eq{ue}, it is
the parameter that connects the spin-down rate of the rigid layer
$\dotOm$ and the mass-loss rate (or the angular momentum loss rate
$\dotJ$ and $\dotM$). Tentatively, we can estimate the product
$\beta\dotM$ (or $\dotJ$) from \eq{beta} assuming high $\beta$. Then,
inequality \eq{condrcap1} leads to

\beq \lambda\frac{3\calN^2R^3}{5GM}<1\; .\eeqn{condrcap}
This new inequality is verified for fast rotating stars where
$\lambda\ll1$ since $\frac{3\calN^2R^3}{5GM}\sim 10$. This means
that the meridional circulation is easily controlled by the
spin-down process in fast rotating stars.

We shall not push the model further because we would clearly need a
model for $\beta$ namely for the relation between mass and angular
momentum losses. We still note that criterion \eq{condrcap}, as criterion
\eq{lamineq1}, does not depend on the mass-loss rate, meaning that once the
shear turbulence due to decelerating layers is settled, the baroclinic
flows is replaced by the spin-down flow provided $\lambda$ is small
enough.

\subsection{Comparison with previous estimates of Zahn (1992)}

We may now compare our estimates of the amplitude of the circulation
with the previous estimate of \cite{zahn92}. From his equation (4.15),
he states that

\[ V_{\rm mer}^{\rm Z92} \approx \frac{15}{8\pi}\frac{\dot{J}}{\rho
R^4\Omega},\]
while our model with the rigid layer and $\beta\gg1$ says that

\[ V_{\rm mer}\simeq\frac{\eps\mathcal{N}^2R}{2\Omega}\csd=
\frac{15\dot{J}}{16\pi\rho\Omega R^4} \; .\]
Hence, up to an unimportant factor 2, the two expressions are
identical. This is also the case for the stress-driven flow. Here,

\[ V_{\rm mer}\sim\frac{\eps\mathcal{N}^2R}{2\Omega}E\tau =
\frac{3\dot{J}}{16\pi\rho\Omega R^4}\]
where the two expressions are also of similar order of magnitude.

\cite{zahn92} distinguishes two regimes for the mass loss: the strong
wind and the moderate wind regimes. For Zahn the strong wind regime
corresponds to an angular momentum loss timescale $t_J$ shorter than
$k^2t_{\rm ES}$, where $k^2$ is such that $k^2MR^2$ is the moment of
inertia of the star, and $t_{\rm ES}$ is the Eddington-Sweet
timescale. Noting that $t_J=J/\dot{J}$ and $t_{\rm ES}=t_{\rm KH}/\eps$,
the strong wind condition $t_J < k^2t_{\rm ES}$ reads

\[ \frac{\dot{J}}{J}>\frac{\eps}{k^2t_{\rm KH}} \orr \dot{M}>\frac{3\eps
M}{2t_{\rm KH}}=\dot{M}_c^{\rm Z92} \]
where $t_{\rm KH}$ is the Kelvin-Helmholtz timescale.
Physically, Zahn's critical mass-loss rate corresponds to the case where a
circulation on the Eddington-Sweet time
scale can no longer supply the wind with enough angular momentum.

This transition may be compared to our mass-loss rate
where turbulence limits the extraction of angular momentum. We may
therefore compare the two rates. We find that

\[ \dot{M}_c^{\rm Z92} =
\frac{3}{\pi}\frac{\eps\calL}{GM\rho\kappa}\lp\frac{\calN}{2\Omega}\rp^2
\dot{M}_{t}\]
where we used $t_{\rm KH}=3GM^2/4\calL R$, $\calL$ being the luminosity
of the star.  A numerical evaluation of the ratio between these two
mass-loss rates gives

\[ \dot{M}_c^{\rm Z92}\sim 30 \dot{M}_t \; .\]
Quite clearly, this mass-loss rate is in the regime of the advection
dominated angular momentum transport as shown by \eq{beta} and our strong
wind limit is lower than Zahn's.

\section{Summary and conclusions}

Investigating the flows induced by mass loss in a rotating star brought
us to a simple model where the star is represented by a ball of (almost)
constant density fluid spun down either by a spinning down outer layer
or by stresses applied to its surface. Both conditions are supplemented
by an outward radial mass flux that mimics the expanding star. Since one open
question about such stars is that of the transport of elements in their
radiative region, the question about mass-losing rotating stars is how
strong should this mass loss be to govern the rotational mixing
that is otherwise triggered by baroclinic flows?

The first step towards an answer to this question is to understand the
underlying fluid dynamics problem. This is the problem of a spin-down
flow and, as the well-known spin-up flow much investigated in the
sixties \cite[][]{green69,Pedlosky79,duck_etal01}, it is a boundary
driven flow. Velocity boundary conditions are unfortunately not well
defined in a star: the spinning down layer, which feels the angular
momentum loss, is part of the star and is likely thickening with time
\cite[][]{LCM96}. If we nevertheless define an interface between the
star and this layer, interface conditions require the continuity of the
velocity field and of the applied stress. In order to make the problem
tractable we considered two idealized cases: In the first case, the
layer is assumed to behave like a rigid shell that spins down and
absorbs matter to feed the wind.  That situation might describe a
turbulent layer threaded by magnetic fields where magnetic fields provide
some rigidity to an outer layer. The other idealized case assumes that
the velocity field might be discontinuous at the interface but that the
stresses are continuous. This condition is inspired by the ocean flows
driven by wind stresses on our Earth.

This latter idealization turned out to be easier to deal with since
it generalizes the results of \cite{R06} in a simple way. It turns out
that the baroclinic flows are overwhelmed by the spin-down flows when
the non-dimensional stress is larger than unity. This simple change is
due to the fact that both of these solutions (baroclinic and spin-down
flows) meet boundary conditions on the stress (namely on the velocity
derivatives).

The second case that we examined has a more complex behaviour. Here the
velocity is prescribed at the interface. By writing the equations in
a frame that co-rotates with the outer spinning-down layer, we can treat,
at linear approximation, the spin-down flow and the baroclinic flow on
an equal footing. When buoyancy can be neglected (needing a fast rotating
star), an analytic solution of the spin-down flow may be derived.The main
result is that as the forcing on spin-down increases, the transition
from the baroclinic flow to the spin-down flow occurs in two steps:
first, the meridional circulation transits to the spin-down circulation
then the differential rotation does the same. The reason for that is that
the baroclinic meridional circulation is of order of the Ekman number $E$
(the non-dimensional measure of viscosity) compared to the associated
differential rotation, while the spin-down meridional circulation is
\od{\sqrt{E}} smaller than its associated differential rotation. Thus,
when the baroclinic meridional circulation is replaced by the spin-down
circulation, its differential rotation is still present.

These fluid dynamics results show that many thresholds might exist in
terms of the spin-down drivings, or in terms of mass-loss rates.

Using the stress prescription, we could identify a transition mass-loss
rate ($\dot{M}_t$) that separates moderate wind regimes where the angular
momentum flux is mainly realized through friction from a strong wind
regime where angular momentum advection dominates.  This peculiar
mass-loss rate is determined by comparing advection of angular momentum
and turbulent stresses that result from \cite{zahn92} prescription
together with the angular velocity profile associated with a constant
specific angular momentum \cite[][]{LCM96}.  It turns out that this
transition mass-loss rate reads

\beq \dot{M}_{\rm t}=\frac{2\pi}{3} \rho\kappa
R\lp\frac{2\Omega}{\calN}\rp^2 \; .
\eeqn{optmlb}
For the two stellar models that we are using as test cases, we find that this
mass-loss rate is $10^{-8}$ and $4\times10^{-7}$ solar mass per year for
a 3\msun\ and 7\msun\ ZAMS stars rotating rapidly (at 200
km/s and 320~km/s resp.). As shown by \eq{optmlb} this mass-loss rate
decreases with rotation as $\Omega^2$.  For winds stronger or equal to
the ``transition'' wind, the spin-down flow completely dominates over
the baroclinic one. When the mass-loss rate decreases below $\dot{M}_t$
one may identify a threshold where the spin-down flow leaves the place
to the baroclinic flows. This critical mass-loss rate is about three
orders of magnitude less than $\dot{M}_t$  for the fast rotating stars
that we are considering. Hence, the model where the spin-down is imposed
via stresses at some boundary display three wind regimes:

\begin{itemize}
\item a weak wind case where the baroclinic flows dominate,

\item a moderate wind case where spin-down flows have superseded
baroclinic flows,

\item a strong wind regime where angular momentum is essentially
advected by the radial outflow.
\end{itemize}

The velocity prescription describing a rigid shell covering the star
interior implies more restrictive conditions for the baroclinic flows
to be superseded by the spin-down flows. The moderate wind threshold
is the same as before, but at this strength the meridional circulation
is the only part of the baroclinic flow that is changed. The mass-loss
rate would need to be increased by a factor $E^{-1/2}$ for spin-down
differential rotation to supersede the baroclinic differential rotation,
but this conclusion is rather uncertain as it requires a modelling of
the relation between mass and angular momentum losses.

We then compared these results with those of \cite{zahn92} when this was
possible. We found that his estimate of the meridional circulation in
the case of a moderate wind well agreed with our estimates either with
our rigid shell model or with the stress-driven spin-down.  On the other
hand, our estimate of the threshold for the strong wind regime is less
by an order of magnitude than that of Zahn's (for our two examples of
stars). We understand this difference with the hypothesis underlying the
two approaches. \cite{zahn92} assumes a slow rotation where circulation
is a transient flow independent of the viscosity, while ours is designed
for fast rotating stars and uses steady solutions where viscosity plays
a crucial role.

The reader may wonder how these results may apply to real stars where
density is far from being constant. Compressibility is certainly one
of the important improvements to make on such a modelling, however,
time-dependence is likely as important especially in the strong wind
regime.  Finally, the interaction between the wind and the star is a
process that requires more investigations. So the numerical estimates
of some remarkable mass-loss rates, although reasonable, should not be
taken at face value in view of the strong hypothesis that lead to them.

The important points of this work is rather the identification of the
various mechanisms that may be at work when the angular momentum loss of a
star is increased.  We hope that this work will be a useful guide in the
understanding of full-numerical multi-dimensional models of mass-losing
rotating stars and that it clearly underscores the crucial points to be
dealt with.

\begin{acknowledgements}
We are very grateful to Sylvie Th\'eado for providing us with stellar
models of ZAMS stars and to Fran\c cois Ligni\`eres for his remarks on
an early version of the manuscript. We also thank the referee, Georges
Meynet for his detailed and constructive criticism of the manuscript. We
acknowledge support of the French Agence Nationale de la Recherche (ANR),
under grant ESTER (ANR-09-BLAN-0140).  This work was also supported by
the Centre National de la Recherche Scientifique (C.N.R.S.), through
the Programme National de Physique Stellaire (P.N.P.S.). The numerical
calculations have been carried out on the CalMip machine of the `Centre
Interuniversitaire de Calcul de Toulouse' (CICT) which is gratefully
acknowledged.

\end{acknowledgements}

\bibliographystyle{aa} 
\bibliography{../../biblio/bibnew} 

\end{document}